\documentclass[aps,prc,letterpaper,11pt,twoside,tightenlines,nofootinbib,showpacs,preprint,superscriptaddress]{revtex4}
\usepackage{amsmath}
\usepackage{graphicx}
\usepackage{subfigure}
\usepackage{color}
\RequirePackage[colorlinks,citecolor=blue,urlcolor=blue,linkcolor=blue]{hyperref}

\newcommand{\Ima}{\textrm{Im}}

\newcommand{\mev}{\textrm{ MeV}}

\begin{document}

\title{Investigations of the $D$-multi-$\rho$ interactions}

\author{C. W. Xiao}
\affiliation{Institut  f\"{u}r Kernphysik (Theorie), Institute for Advanced Simulation, and J\"ulich Center for Hadron Physics, Forschungszentrum J\"ulich, D-52425 J\"{u}lich, Germany}

%\date{\today}

\begin{abstract}

In the present work, which aims at searching for bound sates, the interactions of the $D$-multi-$\rho$ systems are investigated by means of the formalism of the fixed-center-approximation to Faddeev equations. Reproducing the states of $f_2 (1270)$ and $D_1 (2420)$ dynamically in the two-body $\rho\rho$ and $\rho D$ interactions, respectively, as the clusters of the fixed center approximation, the state of $D(3000)^0$ is found as a molecule of $D-f_2$ or $\rho-D_1$ structures in the three-body interactions, where we determine its quantum number $J^P = 2^-$ and find another possible state of $D_2 (3100)$ with isospin $I=3/2$. In our results, there are some other predictions with uncertainties, a $D_3 (3160)$ state with $I(J^P) = \frac{1}{2} (3^+)$ in the four-body interactions, a narrow $D_4 (3730)$ state with $I(J^P) = \frac{1}{2} (4^-)$, a wide $D_4 (3410)$ state of $I(J^P) = \frac{1}{2} (4^-)$, and another wide $D_4 (3770)$ state but with $I(J^P) = \frac{3}{2} (4^-)$ in the five-body interactions, and a $D_5 (3570)$ state with $I(J^P) = \frac{1}{2} (5^+)$ in the six-body interactions. Our results are consistent with the findings of quark models.

\end{abstract}

\maketitle

\section{Introduction}

Multi-body interactions are a topic that has caught much attention in hadronic physics for a long time. Solving the nonrelativistic scattering problems, several sets of equations were suggested in Refs. \cite{Faddeev:1960su,Weinberg:1964zza} for the three-body interactions. Reformulating the Lee model \cite{Lee:1954iq}, Ref. \cite{Rosenberg:1965zz} tried to generalize Faddeev equations \cite{Faddeev:1960su} for the cases of multiparticle scattering. Further, generalized Faddeev equations to a relativistic version were given in Ref. \cite{Alessandrini:1965zz}. But, in practice, it is not easy to evaluate the multiparticale scattering amplitudes by solving these equations, since there are complicated multi-scattering ladder diagrams which should be summed with all of them and a lot of variables involved in the phase space of the integrations. Typically, the difficulties appear at the calculations of the multi-body Green functions which maybe not have ``convenient'' solutions for the multi-scattering amplitudes \cite{Weinberg:1964zza}. In fact, the generalizations of Refs. \cite{Rosenberg:1965zz,Alessandrini:1965zz} avoided these difficulties by redefining the Green functions to simplify the summations. For nuclear physics, these generalizations work much better since the Green functions are dominant by the free Hamiltonian of the three-particle system, where the Faddeev equations can be deduced from the Lippmann-Schwinger equations \cite{Glockle:1970nox}. If no three-body force appearing, the three-body interactions can be derived directly from the Schr\"odinger equations in the Jacobi coordinate space and a bound state of three identical-particle system in a finite volume can be observed \cite{Meissner:2014dea}. Furthermore, without considering the particle spins, Refs. \cite{Elster:1998qv,Bedaque:1998km} had extrapolated the Faddeev equation formalism to the momentum space for the cases of three-boson systems. On the other hand, different with Faddeev equation formulae, Refs. \cite{Gribov:1962fv,Anisovich:1991qk} used a framework of the dispersion relation technique to solve the three-body interactions which try to avoid the problem of ambiguous solutions in Faddeev equations. Also based on the dispersion relation approach, Ref. \cite{Guo:2015kla} discussed the problem in the coupled channel cases for the final state interactions of three spinless-particles. Indeed, evaluating the three-body Green functions both in the coordinate and in the momentum space under the fixed center approximation, the ambiguous results were found in Ref. \cite{Kudryavtsev:2016pzj} where they tried to fix them with some experimental inputs. Besides, driven from an effective field theory for non-relativistic particles, a different formalism was proposed in Ref. \cite{Hammer:2016xye} for the few-body interactions.
 
As discussed above, it is difficult to solve the three-body Faddeev equations (the three-body problem) strictly, and even worse in the relativistic cases. In practical treatments, one should make some assumptions or approximations, and faces with less difficulties in some special cases. For example, the deuteron is a bound state of a pair of nucleons, and it is convenient to let another light particle to collide with them. The hyperon-nucleon interactions, kaons interacted with deuteron, had attracted much attentions for a long time \cite{Toker:1981zh}, which used a non-relativistic Faddeev formalism to study $K^- d \to \pi^- \Lambda p$ reaction with some approximations on the two-body inputs and the restrictions of s-wave interactions. Using chiral perturbation theory, the reaction $\gamma d \to \pi^+ n n$ was accurately calculated up to order $\chi^{5/2}$ in Ref. \cite{Lensky:2005hb} where the three-body dynamics of the $\pi NN$ loop diagrams is discussed in detail. Analogously, Ref. \cite{Miyagawa:2012xz} investigated the reaction $K^- d \to \pi \Sigma n$ only up to second order of the summation of the three-body interaction diagrams, where a three-body unitarity cut was taken to the Green functions. Recently, the work of \cite{Shevchenko:2016wnu} (references therein) made some detail discussions on the three-body antikaon-nucleon interactions and reviewed some results of the antikaon-nucleon systems using the Faddeev-type Alt-Grassberger-Sandhas equations where they applied a free Green function for the three-body propagator (non-relativistic one). Using a different kinematical mechanism compared to Ref. \cite{Miyagawa:2012xz} as clarified in Ref. \cite{Jido:2012cy}, a $K^- d$ quasi-bound state degenerated in spin $S=0$ and $S=1$ was found in Ref. \cite{Bayar:2011qj} as the findings in Ref. \cite{Shevchenko:2016wnu}. Note that, in Ref. \cite{Bayar:2011qj}, they applied a formalism of the fixed-center-approximation (FCA) \cite{Toker:1981zh,Foldy:1945zz,Barrett:1999cw,Kudryavtsev:2012ap} to the Faddeev equations, based on a sets of full Faddeev equations \cite{MartinezTorres:2007sr}. Indeed, a formalism with full Faddeev equations was derived in Ref. \cite{MartinezTorres:2007sr}, which was simplified by an on-shell approximation of the chiral unitary approach and where the three-body loop functions were still complicatedly to evaluate even though the on-shell factorizations had been taken. By taking the FCA to the Faddeev equations, the three-body loop functions (relativistic ones) involved the complicated phase space integrations are absorbed into the form factor of the cluster (fixed center) which simplifies the calculation of the three-body loop functions compared to the ones in the full Faddeev equations \cite{MartinezTorres:2007sr}. Using this formalism, several multi-$\rho(770)$ states, $f_2(1270)$, $\rho_3(1690)$, $f_4(2050)$, $\rho_5(2350)$, and $f_6(2510)$, are dynamically produced in Ref. \cite{Roca:2010tf} and explained as the molecules of an increasing number of $\rho(770)$ particles with parallel spins. An analogous finding was found in Ref. \cite{YamagataSekihara:2010qk} for the $K^*$-multi-$\rho$ systems where the resonances $K^*_2(1430)$, $K^*_3(1780)$, $K^*_4(2045)$, $K^*_5(2380)$ and a new $K^*_6$ were explained as molecules with the components of an increasing number of $\rho(770)$ and one $K^*(892)$ meson. Furthermore, this formalism was extrapolated to investigate the $D^*$-multi-$\rho$ and $K$-multi-$\rho$ interactions in Refs. \cite{Xiao:2012dw,Xiao:2015mqa}. More discussions and applications about this formalism can be found in Ref. \cite{Oset:2015qya} and references therein. In experiments, several $D_J$ and $D_J^*$ states were reported by LHCb \cite{Aaij:2013sza,Aaij:2015sqa} (more discussions about the charm and beauty mesons can be found in the recent review of Ref. \cite{Chen:2016spr}, and some discussions and predictions about $D_{sJ}$ and $B_{sJ}$ can be referred to Ref. \cite{Guo:2011dd}), where the mass of $D^*_3 (2760)$ state was consistent with the predicted one in Ref. \cite{Xiao:2012dw}. Thus, with the motivations from the successes of the FCA to the Faddeev equations and the new findings in the experiments, the present work investigates the interactions of the $D$-multi-$\rho$ systems.

Our paper is organized as follows. We first discuss the formalism of the three-body Faddeev equations with the FCA. Then we show the reproduced results of the two-body interactions. In the following section, the main results for the multi-body interactions are given. We finish with our discussions and conclusions at the end.

\section{Three-body interaction formalism}
\label{secform}

\begin{figure}
\centering
\includegraphics[scale=0.3]{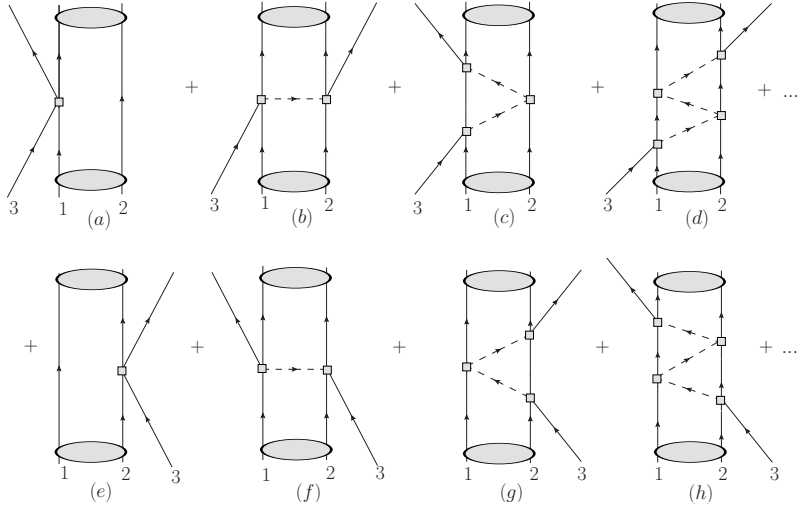}
\caption{Diagrammatic representation of the FCA to Faddeev equations.}\label{fig:FCA}
\end{figure}

In this section, we discuss the formalism of the three-body interactions where how we take the FCA to the Faddeev equations. As suggested in Ref. \cite{Faddeev:1960su}, the total three-body scattering amplitude $T$ can be summed with three partition amplitudes,
\begin{equation}
T=\sum_{i=1}^3 T^{(i)},
\label{eq:faddeev}
\end{equation}
where the component $T^{(i)}$ includes all the possible interactions contributing to the total scattering amplitude $T$ with the particle $i$ being a spectator at the beginning of the interactions. Indeed, the three components of $T^{(i)}$ are identical in their functional form. In fact, they are not independent and correlated with each other, which can be summed with a lot of complicated ladder diagrams when the two-body interactions between three identical particles permute to infinity order. Therefore, the strict solution of Eq. \eqref{eq:faddeev} seems to be impossible for the numerous propagations. Thus, in practice, we should take some approximations to treat the summation. For the cases of having bound states to appear in the two-body subsystems, we can assume the clusters of the bound state as the fixed center of the three-body systems and take the FCA \cite{Toker:1981zh,Foldy:1945zz,Barrett:1999cw,Kudryavtsev:2012ap} to the Faddeev equations as done in Refs. \cite{Roca:2010tf,YamagataSekihara:2010qk}, which can simplify the calculations as discussed in the introduction. For example the cluster coming from $T^{(3)}$, which is formed by the two particles (named as particle 1, 2) and is not much modified by the third particle (particle 3), thus, the $T^{(3)}$ partition amplitude only contributes to the cluster of the FCA, and then the multi-scattering processes just happen at the third particle interacting with the two components of the cluster. Therefore, we can simplify the Faddeev equations of Eq. \eqref{eq:faddeev} as the form
\begin{align}
T_1&=t_1+t_1G_0T_2,\\
T_2&=t_2+t_2G_0T_1,\\
T&=T_1+T_2,
\end{align}
which can be depicted in the diagrams of Fig. \ref{fig:FCA}.
From Fig. \ref{fig:FCA}, the Faddeev equations under the FCA are first a pair of particles (1 and 2) forming a cluster, shown as the grey ellipses, and then particle 3 interacts with the components of the cluster, undergoing all possible multi-scattering processes with those components. Thus, the two partition amplitudes $T_1$ and $T_2$ sum all diagrams of the series of Fig. \ref{fig:FCA} which begin with the interactions of particle 3 with particle 1 of the cluster ($T_1$), or with the particle 2 ($T_2$). The $T_1$ and $T_2$ are the summation of the diagrams in the upper parts and the lower parts, respectively. Finally, the summation of the all of these diagrams are the total three-body scattering amplitude $T$. Besides, the amplitudes $t_1$ and $t_2$ represent the unitary scattering amplitudes for the interactions of particle 3 with particle 1 and 2, respectively, with their coupled channels, which should be taken into account the isospin structures of the subsystems and discussed in details for different cases in Section \ref{secres}. Finally, the function $G_0$ is the propagator of particle 3, given by
\begin{equation}
G_0(s)= \frac{1}{2 M_R} \int \frac{d^3\vec{q}}{(2\pi)^3} F_R(\vec{q}\,) \frac{1}{q^{02}(s)-\vec{q}^{~2}-m_3^2 +i\,\epsilon},
\label{eq:G0s}
\end{equation}
where $q^0 (s)$ is the energy of particle 3 in the three particle system, written
\begin{equation}
q^0(s)=\frac{s+m_3^2-M_R^2}{2\sqrt{s}},
\end{equation}
with $m_3$ is the mass of the third particle and $M_R$ the mass of the cluster, and $F_R(\vec{q}\,)$ is the form factor of the cluster of particles 1 and 2. In our present cases, the expression of the form factors for the $S$-wave bound states is taken as \cite{YamagataSekihara:2010pj},
\begin{align}
\begin{split}
F_R(\vec{q}\,)&=\frac{1}{\mathcal{N}} \int_{|\vec{p}\,|\langle\Lambda', |\vec{p}-\vec{q}\,|\langle\Lambda'} d^3 \vec{p} \; \frac{1}{2 \omega_1(\vec{p}\,)} \frac{1}{2 \omega_2(\vec{p}\,)} \frac{1}{M_R - \omega_1(\vec{p}\,) -\omega_2(\vec{p}\,)} \\
&\quad\frac{1}{2 \omega_1(\vec{p}-\vec{q}\,)} \frac{1}{2 \omega_2(\vec{p}-\vec{q}\,)} \frac{1}{M_R-\omega_1(\vec{p}-\vec{q}\,)-\omega_2(\vec{p}-\vec{q}\,)}, \label{eq:formfactor}
\end{split}\\
\mathcal{N}&=\int_{|\vec{p}\,|\langle\Lambda'} d^3 \vec{p} \; \Big( \frac{1}{2 \omega_1(\vec{p}\,)} \frac{1}{2 \omega_2(\vec{p}\,)} \frac{1}{M_R-\omega_1(\vec{p}\,)- \omega_2(\vec{p}\,)} \Big)^2, \label{eq:formfactorN}
\end{align}
where $\omega_i = \sqrt{\vec{q}\,^2 + m_i^2}$ ($i=1, \, 2$) are the energies of the particles 1, 2, and $m_i$ the corresponding particle masses. We use a cutoff $\Lambda'$ to regularize the integrals of Eqs. \eqref{eq:formfactor} and \eqref{eq:formfactorN}, which is the same one used in the loop functions of the two-body interactions to reproduce the cluster \cite{YamagataSekihara:2010qk}. In fact, in Eq. \eqref{eq:G0s}, there should be also another cutoff, which is taken as $2\,\Lambda'$ for the constraints of the form factor (more discussions will be in Section \ref{secres}). Thus, in our formalism no free parameters are involved.

Next, we should take into account the different weight factors coming from the normalization of the particle fields, where how these factors appear in the single scattering, the double scattering and the total scattering amplitudes \cite{YamagataSekihara:2010qk}. In our present cases, all of the particles, particles 1, 2, 3, and the cluster, are mesons, only related to meson fields. Thus, we can write the $S$ matrix of single scattering diagrams, Figs. \ref{fig:FCA} (a) and (e),
\begin{align}
S^{(1)}_1=&-it_1 (2\pi)^4\,\delta(k+k_R-k'-k'_R) \nonumber \\
&\times \frac{1}{{\cal V}^2} \frac{1}{\sqrt{2\omega_3}} \frac{1}{\sqrt{2\omega'_3}}
 \frac{1}{\sqrt{2\omega_1}} \frac{1}{\sqrt{2\omega'_1}},\label{eq:s11}\\
S^{(1)}_2=&-it_2 (2\pi)^4\,\delta(k+k_R-k'-k'_R)  \nonumber \\
&\times\frac{1}{{\cal V}^2} \frac{1}{\sqrt{2\omega_3}} \frac{1}{\sqrt{2\omega'_3}}
 \frac{1}{\sqrt{2\omega_2}} \frac{1}{\sqrt{2\omega'_2}},\label{eq:s12}
\end{align}
where, $k,\,k'$ ($k_R,\,k'_R$) are the momenta of the initial and final scattering particles ($R$ for the cluster), $\omega_i,\,\omega'_i$ the energies of the initial and final particles,  $\cal V$ is the volume of the box where the states are normalized to unity and the subscripts 1, 2 refer to scattering with particle 1 or 2 of the cluster.

The $S$ matrix of double scattering diagrams, Figs. \ref{fig:FCA} (b) and (f), are given by,
\begin{align}
S^{(2)}=&-i(2\pi)^4 \delta(k+k_R-k'-k'_R) \frac{1}{{\cal V}^2}
\frac{1}{\sqrt{2\omega_3}} \frac{1}{\sqrt{2\omega'_3}}
 \frac{1}{\sqrt{2\omega_1}} \frac{1}{\sqrt{2\omega'_1}}
 \frac{1}{\sqrt{2\omega_2}} \frac{1}{\sqrt{2\omega'_2}} \nonumber \\
&\times\int \frac{d^3q}{(2\pi)^3} F_R(\vec{q}\,) \frac{1}{{q^0}^2-\vec{q}\,^2-m_3^2+i\,\epsilon} t_{1} t_{2},\label{eq:s2}
\end{align}
where $F_R(\vec{q}\,)$ is the cluster form factor that we have discussed above, seen in Eq. \eqref{eq:formfactor}.

Analogously, the full scattering $S$ matrix can be written as,
\begin{equation}
S=-i\, T \, (2\pi)^4 \delta(k+k_R-k'-k'_R)\times\frac{1}{{\cal V}^2}
\frac{1}{\sqrt{2 \omega_3}} \frac{1}{\sqrt{2 \omega'_3}}
\frac{1}{\sqrt{2\omega_R}} \frac{1}{\sqrt{2\omega'_R}}.\label{eq:sful}
\end{equation}
Now, by comparing the different normalization factors of Eqs. \eqref{eq:s11}, \eqref{eq:s12}, \eqref{eq:s2} and \eqref{eq:sful}, we can introduce the weight factors for the elementary amplitudes,
\begin{equation}
\tilde{t_1}=\frac{2M_R}{2m_1}~ t_1,~~~~\tilde{t_2}=\frac{2M_R}{2m_2}~ t_2,
\label{eq:t1t2}
\end{equation}
where we have taken the approximations for the meson fields, $\frac{1}{\sqrt{2 \omega_i}} \simeq \frac{1}{\sqrt{2m_i}}$. One should keep in mind that, when one of the particles in the two-body subsystem (the cluster, the particles 1 and 2) is a baryon, the factors of $2 M_R$ and/or $2 m_i$ in Eqs. \eqref{eq:G0s} and \eqref{eq:t1t2} should be replaced by 1 correspondingly for the approximations of the baryonic fields $\sqrt{\frac{2M_B}{2E_B}} \approx 1$. Finally, the total three-body scattering amplitude $T$ is given by
\begin{equation}
T=T_1+T_2=\frac{\tilde{t_1}+\tilde{t_2}+2~\tilde{t_1}~\tilde{t_2}~G_0}{1-\tilde{t_1}~\tilde{t_2}~G_0^2}. \label{eq:new}
\end{equation}
When $\tilde{t_1}= \tilde{t_2}$ in some cases, it can be simplified as,
\begin{equation}
T=\frac{2 \, \tilde{t_1}}{1-\tilde{t_1} \, G_0}. \label{eq:new2}
\end{equation}
Note that, the FCA to Faddeev equations just can particularly be used to study a three-body system with the subsystem bound or even loose bound, as discussed in Ref. \cite{Bayar:2011qj}. Furthermore, for the unitary amplitudes corresponding to single-scattering contribution, one must take into account the isospin structure of the cluster and write the $t_1$ and $t_2$ ($\tilde{t_1}$ and $\tilde{t_2}$) amplitudes in terms of the isospin amplitudes of the (3,1) and (3,2) systems, discussed later.

From the single scattering $S$ matrix, Eqs.  \eqref{eq:s11} and \eqref{eq:s12}, and the full $S$ matrix, Eq. \eqref{eq:sful}, we should note that the arguments of the amplitudes $T_i(s)$ and $t_i(s_i)$ are different, where $s$ is the total invariant mass of the three-body system, and $s_i$ the invariant mass of the two-body subsystems. The expression of $s_i$ in terms of $s$ is given by \cite{YamagataSekihara:2010qk},
\begin{equation}
s_i=m_3^2+m_i^2+\frac{(M_R^2+m_i^2-m_j^2)(s-m_3^2-M_R^2)}{2M_R^2}, (i,j=1,2,\;i\neq j),
\label{eq:si}
\end{equation}
where the uncertainties of this formula can be referred to the discussions of Ref. \cite{Bayar:2015oea}, and which of cause will introduce some uncertainties to our results as discussed in Ref. \cite{Xiao:2015mqa}.

\section{Basic two-body interaction}
\label{sectwo}

In the former section, we have discussed the formalism of the Faddeev equations under the FCA. Starting with this formalism, first, one should look for bound states in the two-body subsystems which can be treated as the cluster of the fixed center. Then, with this cluster in the fixed center of the three-body system, let the third particle collide with the cluster and interact with two components of the forming cluster. In our present cases, the basic subsystems of the $D$-multi-$\rho$ systems are the two-body $\rho\rho$ and $\rho D$ interactions which were studied in Refs. \cite{Molina:2008jw} and \cite{Gamermann:2007fi}. In their works, the states of $f_2 (1270)$ and $D_1 (2420)$ were dynamically reproduced and associated as the bound states of $\rho\rho$ and $\rho D$ respectively, which are the clusters of the fixed center in our three-body interactions. We briefly summarize their works and reproduce their results, and at the same time obtain the two-body scattering amplitudes of the subsystem which are the essential inputs of the Faddeev equations, seen the discussions in the last section.

\subsection{$\rho\rho$ interaction}

First, we discuss the two-body $\rho\rho$ interactions based on the work of Ref. \cite{Molina:2008jw}, which studied the $\rho\rho$ interactions using the local hidden gauge formalism \cite{Meissner:1987ge,Bando:1987br,Harada:2003jx} and the coupled channel chiral unitary approach \cite{Oller:1997ti,Oset:1997it,Oller:2000fj}. In their work, the $f_2(1270)$ state was dynamically produced in the strong two-body $\rho\rho$ interactions  \footnote{The approximations made in this formalism are being scrutinized in Ref.~\cite{Gulmez:2016scm}.}. From the local hidden gauge Lagrangians, one can derive the potentials of $\rho\rho$ interactions, which are projected to the $s$-wave, for the sectors of spin $S=2$, isospin $I=0$ and $I=2$ obtained,
\begin{eqnarray}
V^{(I=0,S=2)}_{\rho\rho} (s_i) &=& -4 g^2 - 8 g^2 \Big( \frac{3s_i}{4m_\rho^2} - 1 \Big), \\
V^{(I=2,S=2)}_{\rho\rho} (s_i) &=& 2 g^2 + 4 g^2 \Big( \frac{3s_i}{4m_\rho^2} - 1 \Big), 
\end{eqnarray}
where the coupling $g=M_V/(2f_\pi)$, with $M_V$ the vector meson mass and $f_\pi$ the pion decay constant. Using these interaction potentials as the basic inputs, the two-body scattering amplitudes of $\rho\rho$ interactions are given by the on-shell Bethe-Salpeter equations,
\begin{equation}
t^I = [1-V^I G^I]^{-1} V^I,
\label{eq:bse}
\end{equation}
where the kernel $V^I$ is a matrix of the interaction potentials and $G^I$ a diagonal matrix of the loop functions (see Appendix \ref{appa}). Note that the upper index $I$ represents the specific isospin sector for the potentials of the coupled channels sorted by different isospins, which is the isospin structures of the two-body interaction amplitudes.

\begin{figure}
\centering
\includegraphics[width=0.6\textwidth]{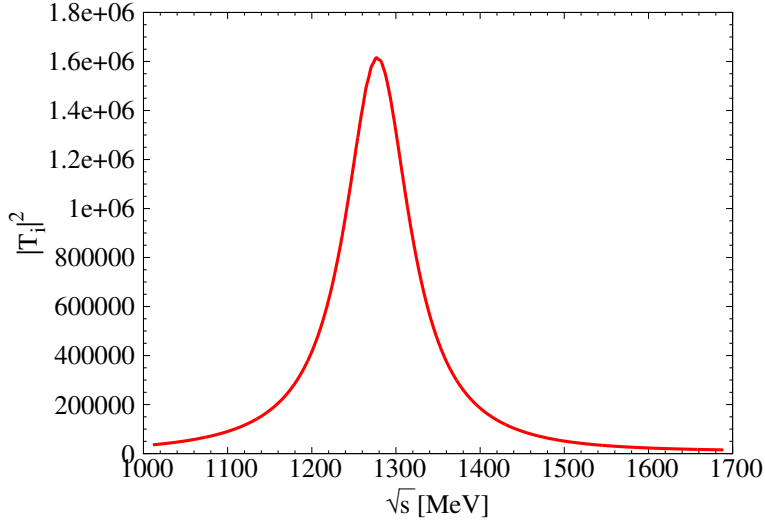}
\caption{Modulus squared of the scattering amplitudes: $|t_{\rho\rho}^{I=0,S=2}|^2$.}\label{fig:trhorho}
\end{figure}

Following the work of Ref. \cite{Molina:2008jw} to dynamically reproduce the $f_2(1270)$ state, we also take into account the contributions of the box diagrams and the $\rho$ mass distributions for its large decay width, where some details can be found in Appendix \ref{appa} and more discussions should be referred to Ref. \cite{Molina:2008jw}. The contributions of the box diagrams were taken into account with two pseudoscalar mesons in the intermediate states of the box, where we just add the imaginary parts of the box diagram contributions for the corrections of the potentials $V^I$ and neglect its real parts, which are much smaller than the imaginary parts. We do not take these intermediate channels in the box diagrams as accounting for the coupled channels since the thresholds of these intermediate channels are much lower than the dominant vector channels. For considering the effects of the large decay width of $\rho$ meson, we do the convolutions for the loop functions in the corresponding channel with the contributions of the $\rho$ mass distributions. Our reproduced results are shown in Fig. \ref{fig:trhorho} for the sector of $I=0,S=2$, which are consistent with Ref. \cite{Molina:2008jw}, and where the structure of the resonance $f_2(1270)$ is shown in the peak of the modulus squared of the amplitudes. Thus, we have dynamically reproduced the $f_2(1270)$ state in the $\rho\rho$ interactions which is one of the clusters in our present work. For the nonresonant amplitude $t_{\rho\rho}^{(I=2,S=2)}$, which is the essential inputs of the two-body interaction amplitudes $t_1, \, t_2$ when we consider the isospin structures of the subsystems (discussed later), we do not show in the figure.

\subsection{$\rho D$ interaction}
\label{subsecRD}

\begin{figure}
\centering
\includegraphics[width=0.6\textwidth]{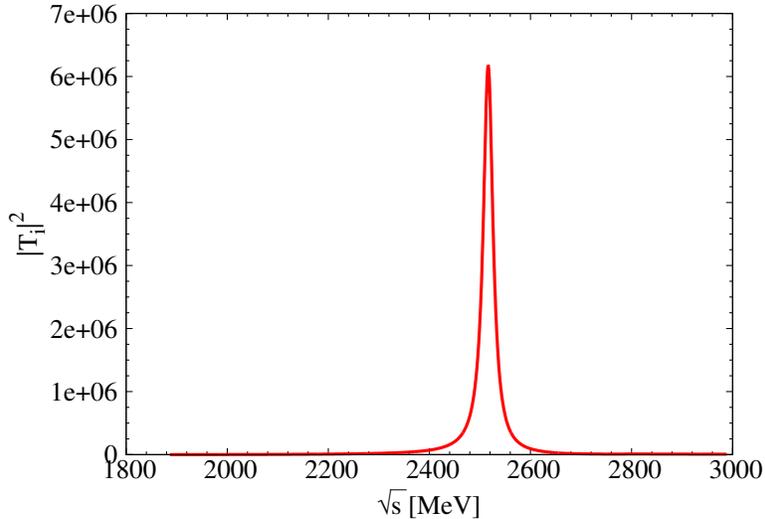}
\caption{Modulus squared of the scattering amplitudes: $|t_{\rho D}^{I=1/2}|^2$.}
\label{fig:trhoD}
\end{figure}

Second, we revisit the $\rho D$ interactions with its couple channels which had been done in Ref. \cite{Gamermann:2007fi}. Following the work of Ref. \cite{Gamermann:2007fi}, there are seven channels coupled to $\rho D$ channel, which are $D^* \pi$, $D^*_s \bar{K}$, $\bar{K}^* D_s$, $D^* \eta$, $\omega D$, $D^* \eta_c$, $J/\psi D$ \footnote{In fact, the thresholds of the $D^* \eta_c$ and $J/\psi D$ channels are far away from the other channels', which have influences negligible on the main results of the bound states in the lower channels and are analogous with the $K \Xi$ channel in the interactions of $\bar{K} N$ and its coupled channels \cite{Oset:1997it}.}. Using the corrected Lagrangian where the masses of the heavy vector mesons are taken into account, the transition potentials are given by
\begin{equation}
V_{ij}^I (s,t,u) = -\frac{C_{ij}^I}{4 f_\pi}(s-u)\epsilon \cdot \epsilon ' ,
\label{eq:vijstu}
\end{equation}
where the lower indexes of $i \, (j)$ channel of the incoming (outgoing) particles, and the coefficients of $C_{ij}^I$ can be found in Appendix A.1 of Ref. \cite{Gamermann:2007fi}, also given in Table. \ref{tab:vijstu} where $C_{ij}^I = C_{ji}^I \; (i \neq j)$ and $\gamma = (m_L / m_H)^2$ with the scales of light and heavy vector mesons, $m_L = 800 \mev$ and $m_H = 2050 \mev$  \cite{Gamermann:2007fi}. 
\begin{table}[htb]
   \renewcommand{\arraystretch}{1.7}
   \setlength{\tabcolsep}{0.4cm}
\centering
\caption{The coefficients of $C_{ij}^I$ in Eq. \eqref{eq:vijstu} for the $I=\frac{1}{2}$ sector.}
\label{tab:vijstu}
\begin{tabular}{ccccccccc}
\hline\hline
Channels & $D^* \pi$ & $\rho\; D$ & $D_s^* \bar{K}$ & $\bar{K}^* D_s$ & $D^* \eta$ & $\omega\; D$ & $D^* \eta_c$ & $J/\psi D$ \\
\hline
$D^* \pi$ 
   & -2 & $\frac{\gamma}{2}$ & $\sqrt{\frac{3}{2}}$ & 0 & 0 & $-\frac{\gamma}{2}$ & 0 & $-\sqrt{2}\;\gamma$ \\
$\rho\; D$  
   &  & -2 & 0 & $-\sqrt{\frac{3}{2}}$ & $\frac{\gamma}{2}$ & 0 & $\sqrt{2}\;\gamma$  & 0 \\
$D_s^* \bar{K}$  
   &  &  & -1 & 0 & $-\sqrt{\frac{3}{2}}$ & $-\sqrt{\frac{2}{3}}\;\gamma$ & 0 &$ \frac{2\gamma}{\sqrt{3}}$ \\
$\bar{K}^* D_s$  
   &  &  &  & -1 & $-\sqrt{\frac{2}{3}}\;\gamma$ & $-\sqrt{\frac{3}{2}}$ & $\frac{2\gamma}{\sqrt{3}}$ & 0 \\
$D^* \eta$  
   &  &  &  &  & 0 & $\frac{\gamma}{6}$ & 0 & $\frac{\sqrt{2}\;\gamma}{3}$ \\
$\omega\; D$  
   &  &  &  &  &  & 0 & $\frac{\sqrt{2}\;\gamma}{3}$ & 0 \\
$D^* \eta_c$   
   &  &  &  &  &  &  & 0 & $\frac{4\gamma}{3}$ \\
$J/\psi D$  
   &  &  &  &  &  &  &  & 0  \\
\hline\hline
\end{tabular}
\end{table}
As done in Ref. \cite{Gamermann:2007fi}, we should do the s-wave projections for the potentials of Eq. \eqref{eq:vijstu}, and then, using the inputs of these potentials for the the on-shell Bethe-Salpeter equations, we can evaluate the scattering amplitudes,
\begin{equation}
t^I  =  [1+V^I \hat{G}^I ]^{-1} (-V^I) \vec{\epsilon} \cdot \vec{\epsilon}\;',
\label{eq:bse2}
\end{equation}
where $\vec{\epsilon}\; (\vec{\epsilon}\;') $ represents a polarization vector of the incoming (outgoing) vector-meson, and now the matrix elements for the loop functions are given by
\begin{equation}
\hat{G}^I_{ii} = (1+ \frac{1}{3}\frac{q_i^2}{M_i^2})G^I_{ii} ,
\end{equation}
with a diagonal matrix of elements $G^I_{ij}$ as the normal one in Eq. \eqref{eq:bse} but using a dimensional regularization expression, as done in Ref. \cite{Gamermann:2007fi}. Note that, Eq. \eqref{eq:bse2} is different from Eq. \eqref{eq:bse} for the factor of $\epsilon \cdot \epsilon '$ in Eq. \eqref{eq:vijstu} where one can refer to Appendix B of Ref. \cite{Roca:2005nm} for more details. Besides, we have checked that, indeed, the term of $\frac{1}{3}\frac{q_i^2}{M_i^2}$ is small and thus it has not much influences of the results, as discussed in Ref. \cite{Roca:2005nm}. Since some of the vector mesons, for example the $\rho$ and $K^*$ mesons, have large decay widths, thus, as the cases of $\rho \rho$ interactions above, we also take into account the mass distributions of these particles, and consider the convolutions of the vector mesons as intermediate state in the loop functions $G^I_{ii}$, where more details can be seen in Appendix \ref{appa}. Furthermore, as done in Ref. \cite{Gamermann:2007fi}, we have ignored the convolutions for the vector mesons with smaller width where the contributions of the mass distributions are trivial. In Fig. \ref{fig:trhoD}, we show the results for the modulus squared of $t_{\rho D}^{I=1/2}$, where the $D_1(2420)$ is reproduced in our work as the cluster of the FCA. We also found three poles in the second Riemann sheets and evaluated the couplings of them to every channels, which are consistent with the results of Ref. \cite{Gamermann:2007fi} and where the conclusions of the second pole corresponding to $D_1(2420)$ state are made. Even though the second pole locates about 100 MeV higher than the mass of the $D_1(2420)$ state, seen Fig. \ref{fig:trhoD} and more discussions referred to Ref. \cite{Gamermann:2007fi}. Besides, the results for the $I=3/2$ sector are not shown in the figure, since there are only two coupled channels, $D^* \pi$ and $\rho D$, and there is no resonance appeared. But this sector should be taken into account for the isospin structures of the subsystems (discussed later), and we also evaluate the amplitude $t_{\rho D}^{I=3/2}$ using the same model.

\section{Results}
\label{secres}

In this section, we show our study results of the $D$-multi-$\rho$ interactions, since we have reproduced the resonances of $f_2(1270)$ and $D_1(2420)$ in the $\rho \rho$ and $\rho D$ two-body interactions in the last section. Assuming the states of $f_2(1270)$ and $D_1(2420)$ as the clusters and using the formalism of Faddeev equations under the FCA as discussed in Sec. \ref{secform}, we can start to investigate the the $D$-multi-$\rho$ interactions. Based on the two options for the clusters in the two-body interactions, the states of $f_2(1270)$ and $D_1(2420)$, thus, for the three-body interactions, we have two possible cases: (i) particle $3=D$, cluster $R=f_2$ (particle $1=\rho,\;2=\rho$) and (ii) $3=\rho$, $R=D_1$ ($1=\rho,\;2=D$). Supposing these two clusters interacting with each other, we can employ the FCA ideas for the four-body interactions, and thus, we also have two cases: (i) $3=f_2$, $R=D_1$ ($1=\rho,\;2=D$) and (ii) $3=D_1$, $R=f_2$ ($1=\rho,\;2=\rho$). 
If we can find some bound states in the four-body interactions, we can treat these bound states as new cluster of the FCA and let another meson collide with them. Therefore, we can extrapolate the FCA formalism to the five-body interactions. We have known the state $f_4$ explained as the molecules of four $\rho$ mesons \cite{Roca:2010tf}, and assumed a new bound state, $D_3$, found in our four-body interactions, and thus, there are also two cases for the five-body interactions, (i) $3=D$, $R=f_4$ ($1=f_2,\;2=f_2$) and (ii) $3=\rho$, $R=D_3$ ($1=f_2,\;2=D_1$). Following, for the six-body interactions, let the two-body cluster, $f_2(1270)$ or $D_1(2420)$, collide with the four-body bound states, $f_4$ or $D_3$, we have two options as well: (i) $3=D_1$, $R=f_4$ ($1=f_2,\;2=f_2$) and (ii) $3=f_2$, $R=D_3$ ($1=f_2,\;2=D_1$). Finally, we summarize all the possible cases for the $D$-multi-$\rho$ interactions in Table \ref{tab:cases}, where we only consider up to number six (n-body) for the maximum binding energy in per $\rho$ meson evaluated in Ref. \cite{Roca:2010tf}. Next, our results for all these cases are discussed below.
\begin{table}[htb]
   \setlength{\tabcolsep}{0.4cm}
\centering
\caption{All possible cases for the $D$-multi-$\rho$ interactions.}
\label{tab:cases}
\begin{tabular}{cccc}
\hline\hline
Particles: &  3  &  Cluster (1,2)  &   Amplitudes  \\
\hline
Two-body
  & $\rho$ & $D$ &  $t_{\rho D}$  \\
  & $\rho$ & $\rho$ & $t_{\rho\rho}$ \\
\hline
Three-body
  & $D$ & $f_2\;(\rho\rho)$ & $T_{D-f_2}$  \\
  & $\rho$ & $D_1\;(\rho D)$ & $T_{\rho-D_1}$  \\
\hline
Four-body
  & $D_1$ & $f_2\;(\rho\rho)$ & $T_{D_1-f_2}$   \\
  & $f_2$ & $D_1\;(\rho D)$ & $T_{f_2-D_1}$ \\
\hline
Five-body
  & $D$ & $f_4\;(f_2 f_2)$ & $T_{D-f_4}$  \\
  & $\rho$ & $D_3\;(f_2 D_1)$ & $T_{\rho-D_3}$   \\
\hline
Six-body
  & $D_1$ & $f_4\;(f_2 f_2)$ & $T_{D_1-f_4}$  \\
  & $f_2$ & $D_3\;(f_2 D_1)$ & $T_{f_2-D_3}$  \\
\hline\hline
\end{tabular}
\end{table}

\subsection{Three-body interactions}
\label{threeb}

\begin{figure}
\centering
\includegraphics[scale=0.8]{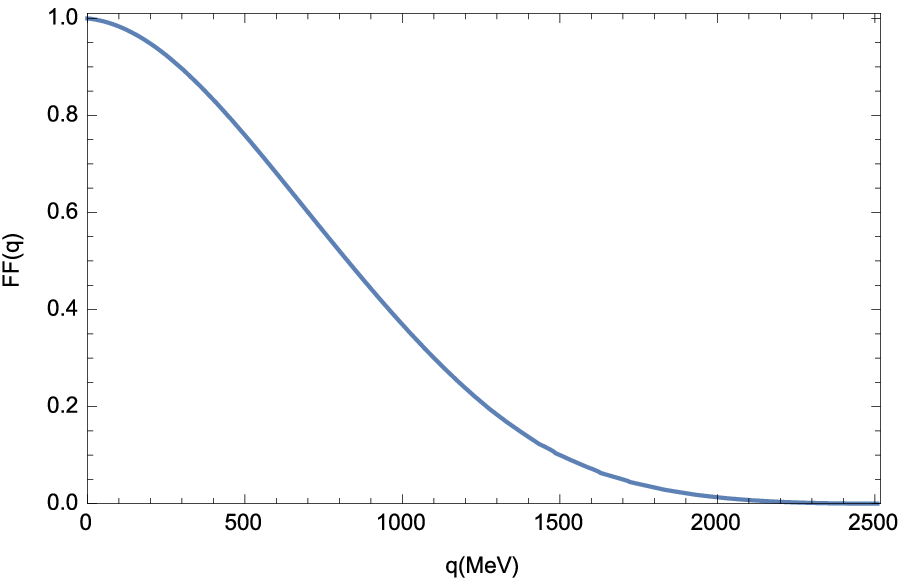}
\includegraphics[scale=0.8]{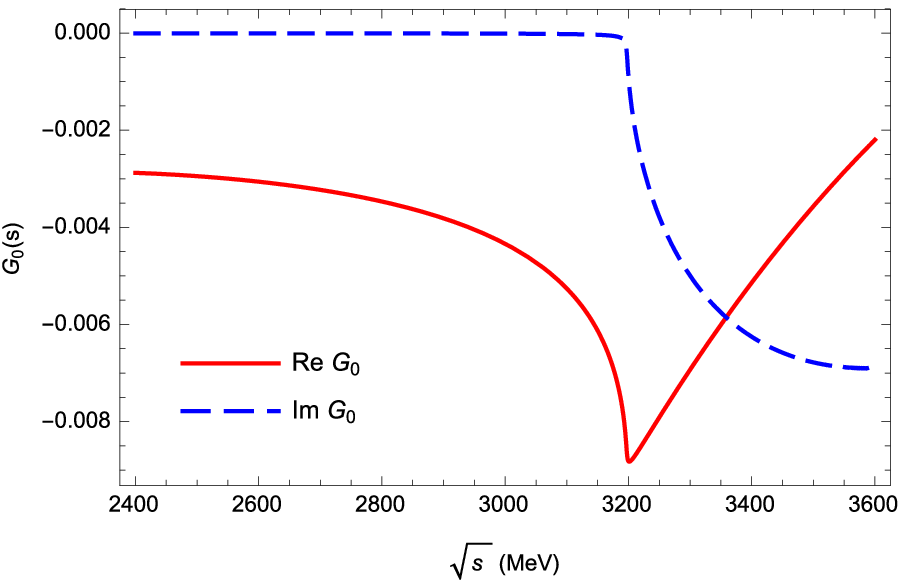}
\caption{The results of the form factor of $D_1(2420)$ state (left) and the real and imaginary parts of the $G_0$ function in $\rho-D_1$ scatterings (right).} \label{fig:t3ffg}
\end{figure}

First, we begin with the discussions for the three-body interactions. From Table \ref{tab:cases}, there are two possible structures, $D-f_2(\rho\rho)$ and $\rho-D_1(\rho D)$: (i) $3=D$, $R=f_2$ ($1=\rho,\;2=\rho$) and (ii) $3=\rho$, $R=D_1$ ($1=\rho,\;2=D$). As the basic inputs of the FCA to Faddeev equations, seen Sec. \ref{secform}, the $t_1$ and $t_2$ amplitudes of the (3,1) and (3,2) subsystems, $t_1 = t_2 = t_{\rho D}$ for $D-f_2(\rho\rho)$ scatterings and $t_1 = t_{\rho\rho},\; t_2=t_{\rho D}$ for $\rho-D_1(\rho D)$ scatterings, have been discussed in the last section, following Refs. \cite{Molina:2008jw,Gamermann:2007fi}. But, to evaluate the form factor of the cluster, seen Eq. \eqref{eq:formfactor}, a cutoff $\Lambda'$ should be used, which is the same as the cutoff $q_{max}$ applied in the loop functions of the two-body interactions, discussed in Sec. \ref{secform}. Note that, the dimensional regularization scheme is used for the loop functions in Ref. \cite{Gamermann:2007fi}, seen in Appendix \ref{appa}. Following the method mentioned in Ref. \cite{Xiao:2011rc}, we can match two regularization schemes of the loop functions at the threshold of the corresponding channels, and then, the equivalent parameters are obtained. Therefore, we obtain $q_{max}=1254\mev$ for the $D_1(2420)$ cluster. Besides, we take $q_{max}=875\mev$ for the $f_2(1270)$ cluster as the one used in Ref. \cite{Molina:2008jw}. Indeed, we do not introduce any free parameter as discussed before. In Fig. \ref{fig:t3ffg}, we show the results of the form factor of $D_1(2420)$ cluster and the $G_0$ function of $\rho-D_1(\rho D)$ scatterings. Form the results of the form factor in the left panel of Fig. \ref{fig:t3ffg}, one can find that a cutoff $2\,\Lambda'$ is enough for the $G_0$ function of Eq. \eqref{eq:G0s}, as discussed in Sec. \ref{secform} and seen the results on the right panel of Fig. \ref{fig:t3ffg} for  the real and imaginary parts of the $G_0$.

As mentioned in the formalism of Sec. \ref{secform}, the isospin structures of the subsystems for the two-body amplitudes $t_1$ and $t_2$ should be taken into account. In the case of $D-f_2(\rho\rho)$ scatterings, since the cluster of $f_2$ resonance has isospin $I=0$, the two $\rho$ mesons combine in an $I=0$ state, having
\begin{equation} 
|\rho\rho\rangle^{(0,0)}=\frac{1}{\sqrt{3}} \big[ |(1,-1)\rangle \;+\; |(-1,1)\rangle \;-\; |(0,0)\rangle \big],
\end{equation}
where $|(1,-1)\rangle$ denotes $|(I_z^1,I_z^2)\rangle$ with the isospin $I_z$ (third) components of particles 1 and 2, and $|\rho\rho\rangle^{(0,0)}$ means $|\rho\rho\rangle^{(I,I_z)}$. Then, taken the third particle of $D$ meson as $|I_z^3\rangle=|\frac{1}{2}\rangle$, we obtain
\begin{equation} 
\begin{split}
T^{(\frac{1}{2},\frac{1}{2})}_{D-f_2}=&(\langle D|^{(\frac{1}{2},\frac{1}{2})} \otimes \langle\rho\rho|^{(0,0)})\,(\hat{t}_{31}+\hat{t}_{32})\,(|D\rangle^{(\frac{1}{2},\frac{1}{2})} \otimes |\rho\rho\rangle^{(0,0)})\\
=&\Big[ \langle\frac{1}{2}| \otimes \frac{1}{\sqrt{3}} \Big( \langle(1,-1)| + \langle(-1,1)| - \langle(0,0)| \Big) \Big]\,(\hat{t}_{31}+\hat{t}_{32})\,\Big[ |\frac{1}{2}\rangle \\
&\otimes \frac{1}{\sqrt{3}} \Big( |(1,-1)\rangle + |(-1,1)\rangle - |(0,0)\rangle \Big) \Big]\\
=&\frac{1}{3} \Big[ \langle(\frac{3}{2},\frac{3}{2}),-1| + \sqrt{\frac{1}{3}} \langle(\frac{3}{2},-\frac{1}{2}),1| + \sqrt{\frac{2}{3}} \langle(\frac{1}{2},-\frac{1}{2}),1| - \sqrt{\frac{2}{3}} \langle(\frac{3}{2},\frac{1}{2}),0| \\
&- \sqrt{\frac{1}{3}} \langle(\frac{1}{2},\frac{1}{2}),0| \Big] \hat{t}_{31} \Big[ |(\frac{3}{2},\frac{3}{2}),-1\rangle + \sqrt{\frac{1}{3}} |(\frac{3}{2},-\frac{1}{2}),1\rangle + \sqrt{\frac{2}{3}} |(\frac{1}{2},-\frac{1}{2}),1\rangle \\
&- \sqrt{\frac{2}{3}} |(\frac{3}{2},\frac{1}{2}),0\rangle - \sqrt{\frac{1}{3}} |(\frac{1}{2},\frac{1}{2}),0\rangle \Big] + \cdots,
\label{eq:tDf2}
\end{split}
\end{equation}
where we take the notation of $|(\frac{3}{2},\frac{3}{2}),-1\rangle \equiv |(I^{31},I_z^{31}),I_z^2\rangle$ for $t_{31}$, and the analogous derivations for $t_{32}$ are not shown. Finally, the amplitudes considering the isospin structures are given by
\begin{equation}
t_1 = t_{\rho D} = \frac{1}{3} \big(2 \;t_{31}^{I=3/2} + t_{31}^{I=1/2} \big),\quad t_2 = t_1, 
\label{eq:tDrho}
\end{equation}
where the amplitudes $t_{31}^{I=1/2}$ and $t_{31}^{I=3/2}$ for $\rho D$ scatterings have been evaluated in Sec. \ref{sectwo}.

But, in the case of $\rho-D_1(\rho D)$, the isospin structure relationships are a bit complicated, since the isospins of $\rho$ and $D_1$ are $I_\rho=1$ and $I_{D_1}=\frac{1}{2}$ which lead to the total isospin of the three-body system having two cases: $I_{total}=\frac{1}{2}$ and $I_{total}=\frac{3}{2}$. Therefore, performing a similar derivation of Eq. \eqref{eq:tDf2}, we obtain
\begin{equation}
\begin{split}
&T_{\rho-D_1}^{(I=1/2)}: \quad t_1 = t_{\rho\rho} = \frac{2}{3} \;t_{31}^{I=0}, \quad t_2 = t_{\rho D} = \frac{1}{9} \big( 8 \;t_{32}^{I=3/2} + t_{32}^{I=1/2} \big); \\
&T_{\rho-D_1}^{(I=3/2)}: \quad t_1 = t_{\rho\rho} = \frac{5}{6} \;t_{31}^{I=2}, \quad t_2 = t_{\rho D} = \frac{1}{9} \big( 5 \;t_{32}^{I=3/2} + 4 \;t_{32}^{I=1/2} \big);
\end{split}
\end{equation}
where the amplitudes $t_{31}^{I=0}$, $t_{31}^{I=2}$ are the ones for the $\rho \rho$ interactions, and $t_{32}^{I=1/2}$, $t_{32}^{I=3/2}$  for the $\rho D$ interactions, are given in the last section as well.

\begin{figure}
\centering
\includegraphics[scale=0.6]{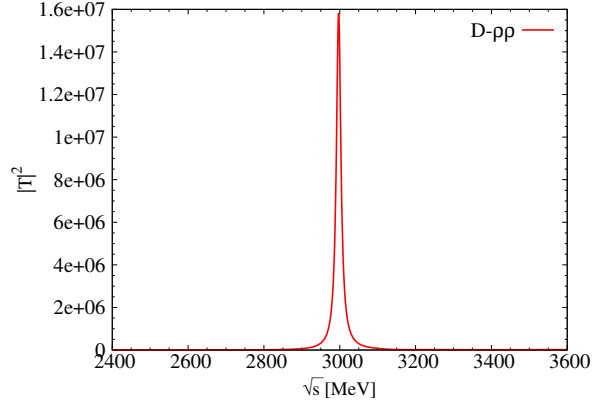}
\caption{Modulus squared of the amplitudes $T_{D-f2}$.}
\label{fig:t3Drhorho}
\end{figure}

\begin{figure}
\centering
\includegraphics[scale=0.6]{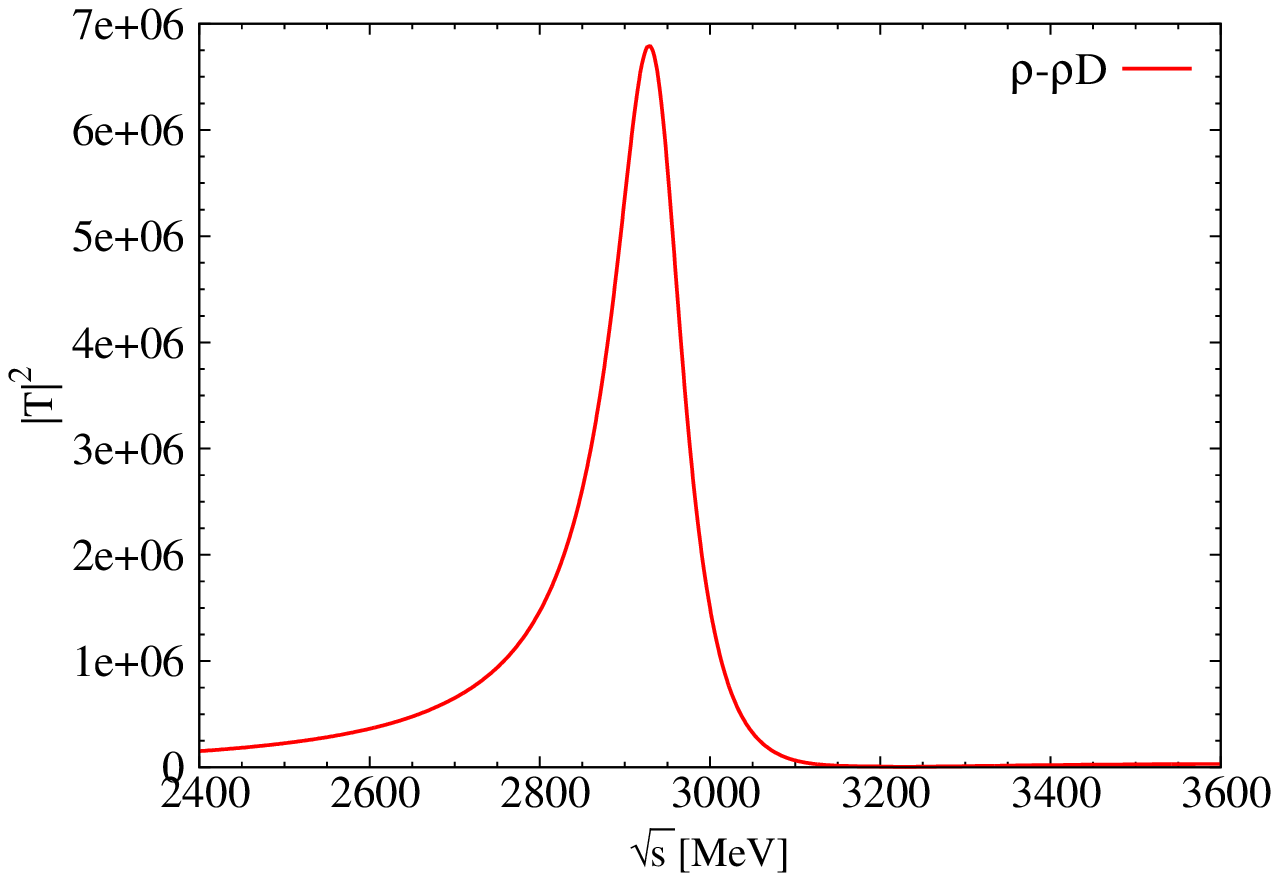}
\includegraphics[scale=0.6]{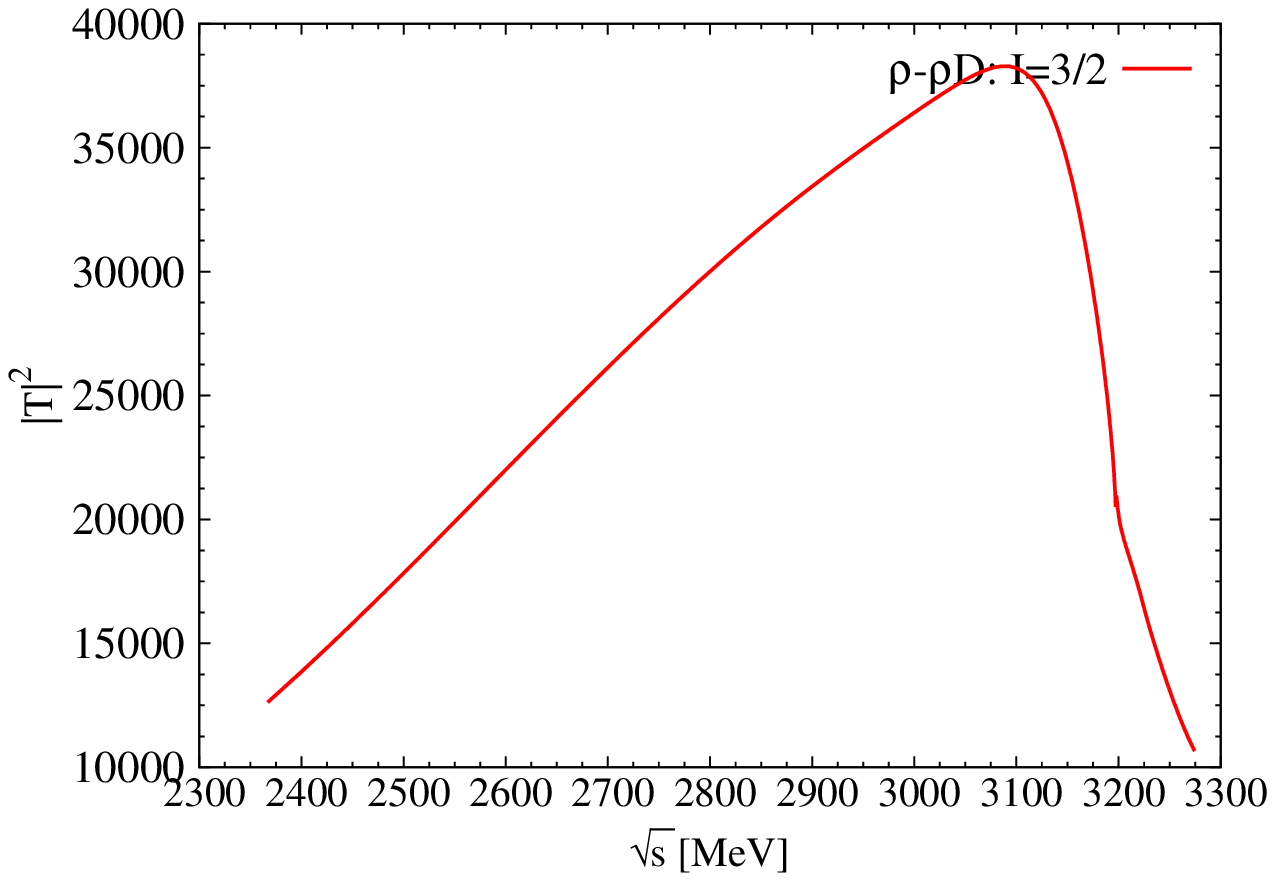}
\caption{Modulus squared of the amplitudes $T_{\rho-D_1}$. Left: $I_{total}=\frac{1}{2}$; Right: $I_{total}=\frac{3}{2}$.} \label{fig:t3rhorhoD}
\end{figure}

Using these two-body inputs for the FCA to the Faddeev equations, the three-body scattering amplitudes can be evaluated. For the first case of $D-f_2(\rho\rho)$ scatterings, the results of the modulus squared of the amplitudes for $|T_{D-f_2}^{I=1/2}|^2$ are shown in Fig. \ref{fig:t3Drhorho}, where a narrow peak around $2997\mev$ with a width of about $14\mev$ is found. This peak is about 150 MeV below the $D-f_2$ threshold. For the second case of $\rho-D_1(\rho D)$ interactions, we show the results in Fig. \ref{fig:t3rhorhoD} for the amplitudes of $|T_{\rho-D_1}^{I=1/2}|^2$ (left) and $|T_{\rho-D_1}^{I=3/2}|^2$ (right). From the left panel of Fig. \ref{fig:t3rhorhoD} for $|T_{\rho-D_1}^{I=1/2}|^2$, we find a clear peak around the energy $2929\mev$ with a width about $103\mev$, which is about $270\mev$ below the $\rho-D_1$ threshold. From Sec. \ref{sectwo}, we can see that both of $\rho \rho$ and $\rho D$ are strongly bound to form the $f_2$ and $D_1$ states. Thus, the large bindings in these peaks will be acceptable. In Particle Data Group (PDG) \cite{pdg2016}, there is a $D(3000)^0$ state with isospin $I=1/2$ and unknown $J^P$ quantum numbers in the lists, since both natural- and unnatural-parity components are observed in LHCb experiments \cite{Aaij:2013sza}, where it was reported as a $D_J (3000)$ state with the mass $(2971.8 \pm 8.7)\mev$ and the width $(188.1 \pm 44.8)\mev$ seen in the $D^* \pi$ invariant mass spectrum, and a $D_J^* (3000)$ state with the mass $(3008.1 \pm 4.0)\mev$ and the width $(110.5 \pm 11.5)\mev$ found in the $D \pi$ invariant mass spectrum. The findings of the $D_J^{(*)} (3000)$ states had caught much theoretical attentions, where some theoretical interpretations for them were made from the heavy meson effective theory \cite{Wang:2013tka,Batra:2015cua}, the relativistic quark model \cite{Sun:2013qca,Godfrey:2015dva,Liu:2016efm,Segovia:2013wma,Liu:2015uya}, the $^3P_0$ decay model \cite{Yu:2014dda,Lu:2014zua}, the chiral quark model \cite{Xiao:2014ura}, and the relativistic potential model with the heavy quark symmetry \cite{Matsuki:2016hzk}. In quark models, the $D_J^* (3000)$ state was mostly assigned as a $2^3P_0$ state with a predicted mass of about $2932\mev$ \cite{Sun:2013qca,Godfrey:2015dva,Lu:2014zua,Matsuki:2016hzk}, and the $D_J (3000)$ state mostly as a $3^1S_0$ state with a mass about $3068\mev$ \cite{Sun:2013qca,Godfrey:2015dva,Lu:2014zua}. But, in Ref. \cite{Xiao:2014ura}, the theoretical mass was about $2949/2919\mev$ for $2^3P_0$ state, and $2995/2932\mev$ for $2P_1$ state which was associated to $D_J (3000)$ state as concluded in Ref. \cite{Sun:2013qca}. On the other hand, there were different assignments in Ref. \cite{Liu:2016efm}, but, the predicted masses were consistent with the other quark models, $2928\mev$ for the $D_J (3000)$ state and $2957\mev$ for the $D_J^* (3000)$ state. More possible assignments for the $D_J^{(*)} (3000)$ states and more discussions can also be found in the recent review of Ref. \cite{Chen:2016spr}. In the present work, our predicted masses both in the $D-f_2(\rho\rho)$ and the $\rho-D_1(\rho D)$ scatterings are consistent with the $D(3000)^0$ state in PDG and also consistent with the results of the other models mentioned above, even though the widths of our model are a bit smaller since we do not take into account the contributions of the large width of the clusters, the $f_2$ and $D_1$ states, seen Ref. \cite{Xiao:2015mqa} for more discussions. Thus, in our model, we can conclude that the $D(3000)^0$ state can be a molecular state of $D-f_2$ or $\rho-D_1$ structures with some uncertainties, where one can determine its $J^P = 2^-$. Our predictions for $J^P$ quantum numbers are consistent with the one of possible assignments in Ref. \cite{Batra:2015cua}. From the results of $|T_{\rho-D_1}^{I=3/2}|^2$ on the right panel of Fig. \ref{fig:t3rhorhoD}, it seems that there is a resonant structure located at about $3100\mev$ but with much larger width (about $650\mev$), which is about 100 MeV lower than the $\rho-D_1$ threshold. The strength of $|T_{\rho-D_1}^{I=3/2}|^2$ is two magnitudes smaller than $|T_{\rho-D_1}^{I=1/2}|^2$. Therefore, this resonant structure has much larger uncertainties.

\subsection{Four-body interactions}
\label{fourb}

For the four-body interactions, there are also two possible cases: (i) particle $3=f_2$, cluster $R=D_1$ ($1=\rho,\;2=D$), or (ii) particle $3=D_1$, resonance $R=f_2$ ($1=\rho,\;2=\rho$). Due to the isospins of the two clusters, $I_{f_2} = 0$ and $I_{D_1} = \frac{1}{2}$, the total isospins of two four-body systems are $I_{total} = \frac{1}{2}$. Let's discuss the basic inputs of the FCA formalism for the four-body interactions. For the first case of $f_2$ colliding with the $D_1$, the amplitudes $t_1 = t_{f_2 \rho} = T_{\rho-f_2}$, which can be reproduced by following Ref. \cite{Roca:2010tf}, and $t_2 = t_{f_2 D} = T_{D-f_2}$ has been calculated in the former subsection. For the second case of $D_1$ interacting with the $f_2$, the amplitudes $t_1 = t_2 = t_{D_1 \rho} = T_{\rho-D_1}$ are also obtained in the former subsection. Since the isospins of $D_1$ and $D$ are $I=\frac{1}{2}$, the isospin structure for the input amplitues of $D_1-f_2$ is similar to the case of $D-f_2$. Thus, using Eq. \eqref{eq:tDrho} we obtain
\begin{equation}
t_1 = T_{\rho D_1} = \frac{1}{3} \big( 2 T_{31}^{I=3/2} + T_{31}^{I=1/2} \big),\quad t_2 = t_1,
\label{eq:threet3}
\end{equation}
where the isospin amplitudes of $T_{\rho D_1}$ can be evaluated from the last subsection.

\begin{figure}
\centering
\includegraphics[scale=0.6]{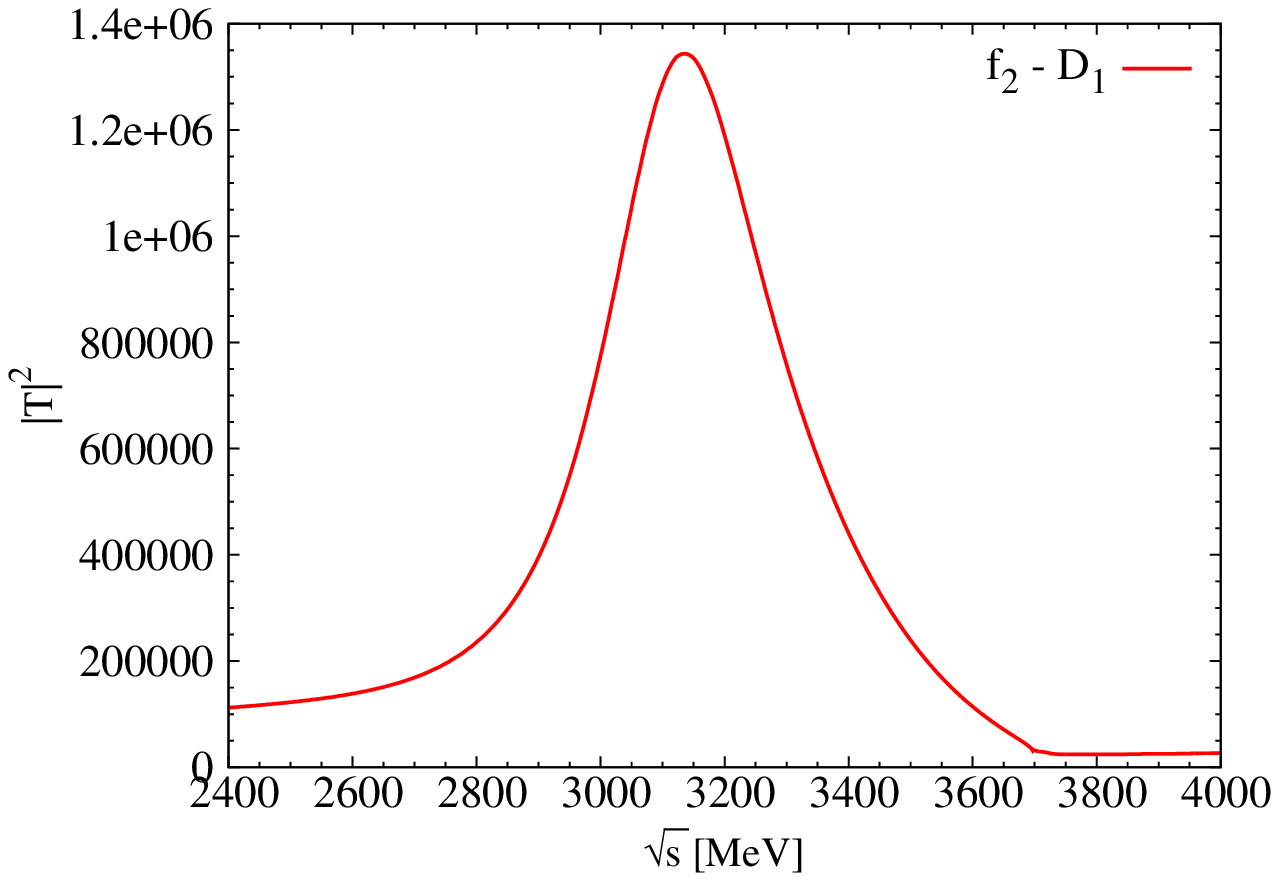}
\includegraphics[scale=0.6]{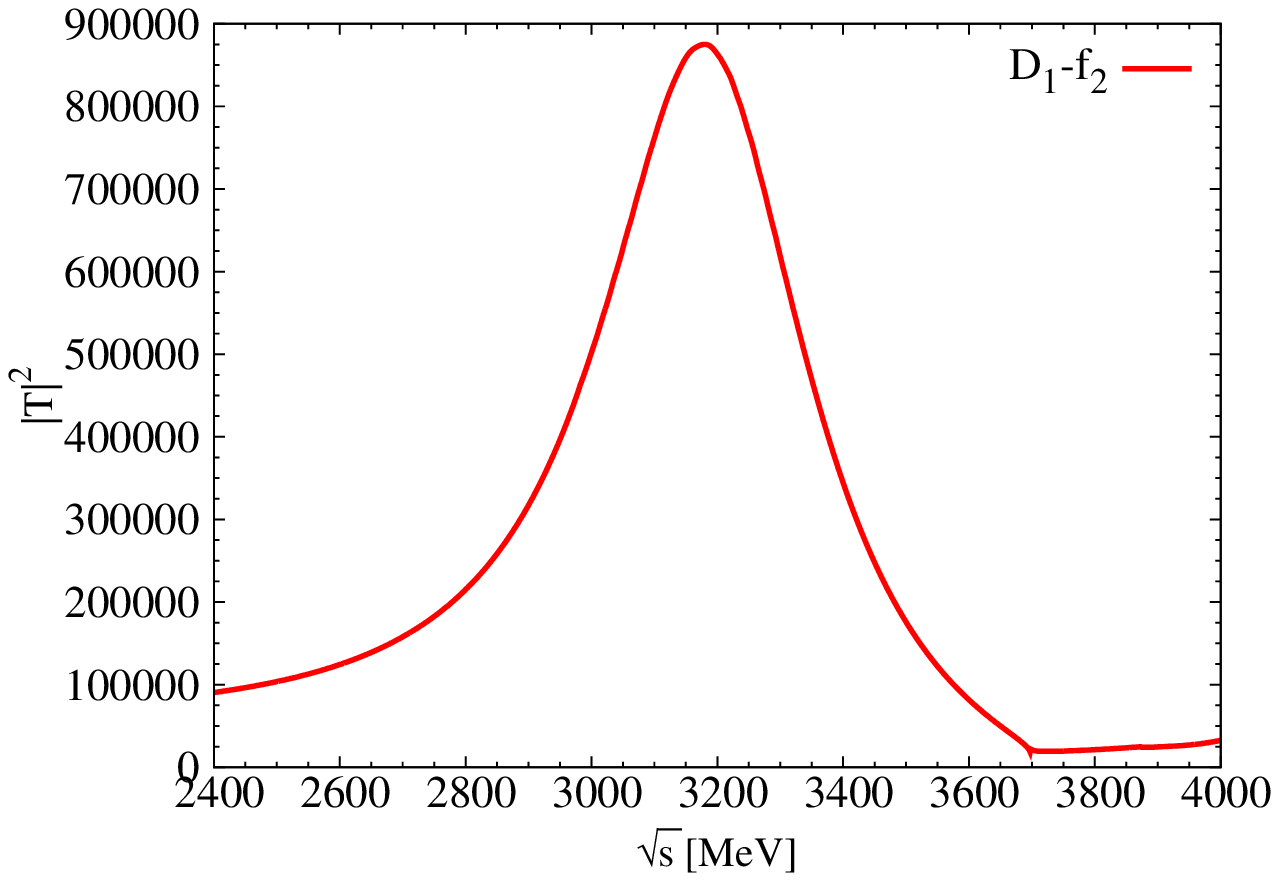}
\caption{Modulus squared of the scattering amplitudes: $T_{f_2-D_1}$ (left) and $T_{D_1-f2}$ (right).}
\label{fig:t4D1f2}
\end{figure}

In Fig. \ref{fig:t4D1f2}, we show our results of the modulus squared of the scattering amplitudes, where $|T_{f_2-D_1}^{I=1/2}|^2$ is on the left and $|T_{D_1-f_2}^{I=1/2}|^2$ on the right. In the left panel of Fig. \ref{fig:t4D1f2}, one can see a resonant peak at the region of about $3135\mev$ with a width about $344\mev$. In the right panel of Fig. \ref{fig:t4D1f2}, we also find a resonant structure around the energy $3180\mev$, of which the width is about $390\mev$. The strengths of both peaks of $|T_{D_1-f_2}^{I=1/2}|^2$ and $|T_{f_2-D_1}^{I=1/2}|^2$ are nearly the same magnitudes. For these two resonant structures, there is no corresponding state listed in the PDG \cite{pdg2016}, since their masses are almost out of the searching range of the experiments  \cite{Aaij:2013sza}. There were also some predicted charm states with masses around the region of $2816\sim 3288\mev$ in the quark models \cite{Sun:2013qca,Godfrey:2015dva,Liu:2016efm,Lu:2014zua,Matsuki:2016hzk}, which are consistent with our results. Therefore, a new $D_3 (3160)$ state is predicted in our model with some uncertainties, of which the mass is about $3135 \sim 3190\mev$ and the width about $344 \sim 390\mev$.

\subsection{Five-body interactions}

In the former subsection, we have shown the results for the four-body interactions, where a new $D_3$ state is found as we expected. Therefore, this predicted $D_3$ state also can be a cluster of the FCA. Thus, there also are two options for the clusters in the five-body interactions, as assuming in Table. \ref{tab:cases}, one of which is the $f_4$ state listed in the PDG and studied in Ref. \cite{Roca:2010tf}, and the other one the bound state of $D_3$ predicted above. Then, letting another particle ($D$ or $\rho$) collide with them, we have two possibilities: (i) particle $3=D$, cluster $R=f_4$ ($1=f_2,\;2=f_2$), or (ii) $3=\rho$, $R=D_3$ ($1=f_2,\;2=D_1$). For the first option, since the isospin $I_{f_4} = 0$ and $I_{D_3} = \frac{1}{2}$, the total isospin of the $D-f_4$ system is only one possibility, $I_{total} = \frac{1}{2}$, of which the isospin structure is similar to the three-body interactions of $D$ particle colliding with $f_2$ before ($D-f_2$). The input amplitudes $t_1 = t_2 = t_{D f_2} = T_{D-f_2}^{(I=1/2)}$ have been evaluated in Subsec. \ref{threeb} of the three-body interactions. But for the second option, the total isospin of $\rho-D_3$ system has two probabilities, $I_{total} = \frac{1}{2}$ or $I_{total} = \frac{3}{2}$, where the situations of $\rho-D_3$ are analogous to the one of the  three-body interactions $\rho-D_1$ in Subsec. \ref{threeb}. Thus, doing a similar derivation as Eq. \eqref{eq:tDf2}, we obtain the isospin structures for the case of $\rho-D_3$ interactions,
\begin{equation} 
\begin{split}
&T_{\rho-D_3}^{(I=1/2)}: \quad t_1 = t_{\rho f_2} = T_{31}^{(I=1)}, \quad t_2 = t_{\rho D_1} =  T_{32}^{I=1/2}; \\ 
&T_{\rho-D_3}^{(I=3/2)}: \quad t_1 = t_{\rho f_2} = T_{31}^{(I=1)}, \quad t_2 = t_{\rho D_1} =  T_{32}^{I=3/2},
\end{split}
\end{equation}
where the amplitude $T_{31}^{(I=1)}$ for $T_{\rho-f_2}$ has been calculated in the Subsec. \ref{fourb} following Ref. \cite{Roca:2010tf}, and the amplitudes $T_{32}^{I=1/2}$ and $T_{32}^{I=3/2}$, for $T_{\rho-D_1}^{I=1/2}$ and $T_{\rho-D_1}^{I=3/2}$, are computed in Subsec. \ref{threeb}.

\begin{figure}
\centering
\includegraphics[scale=0.6]{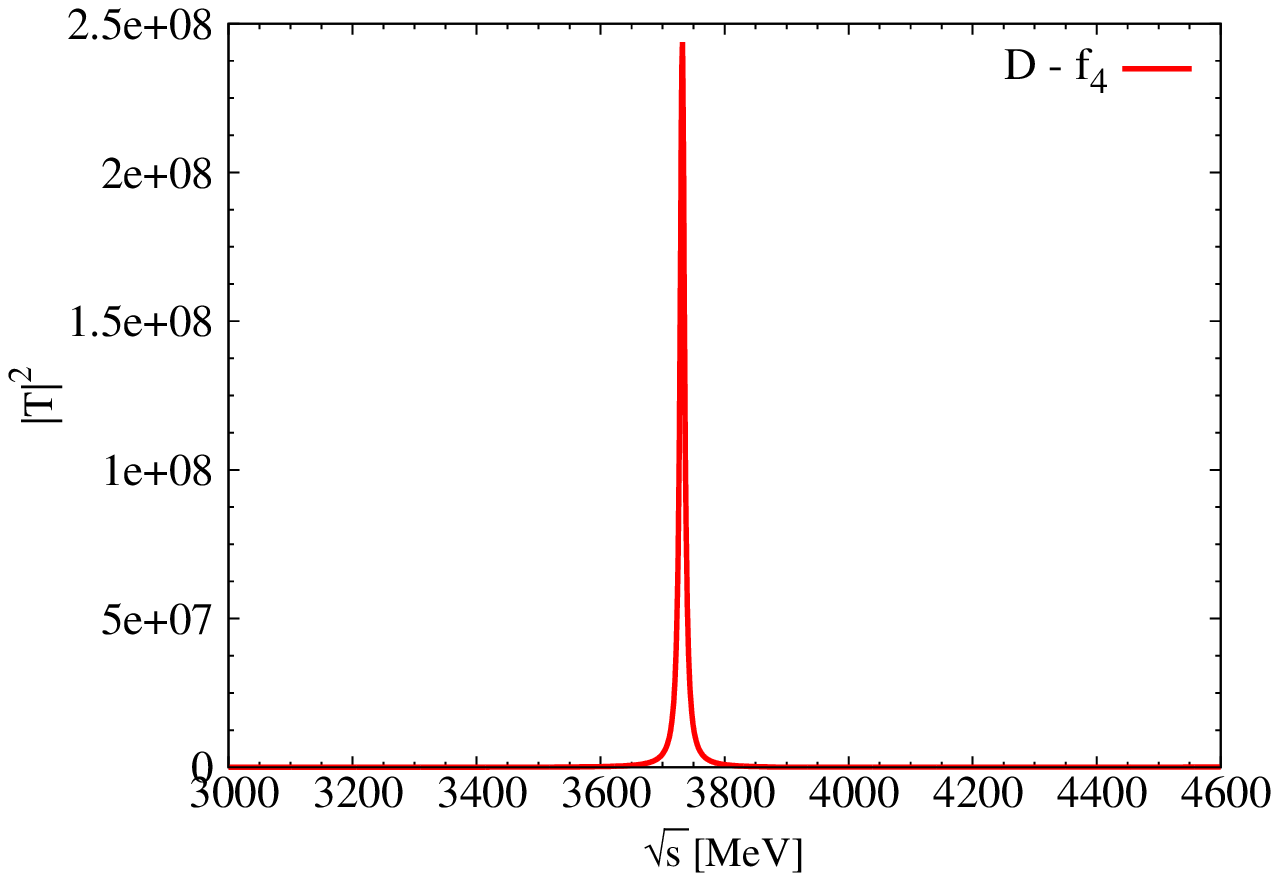}
\includegraphics[scale=0.6]{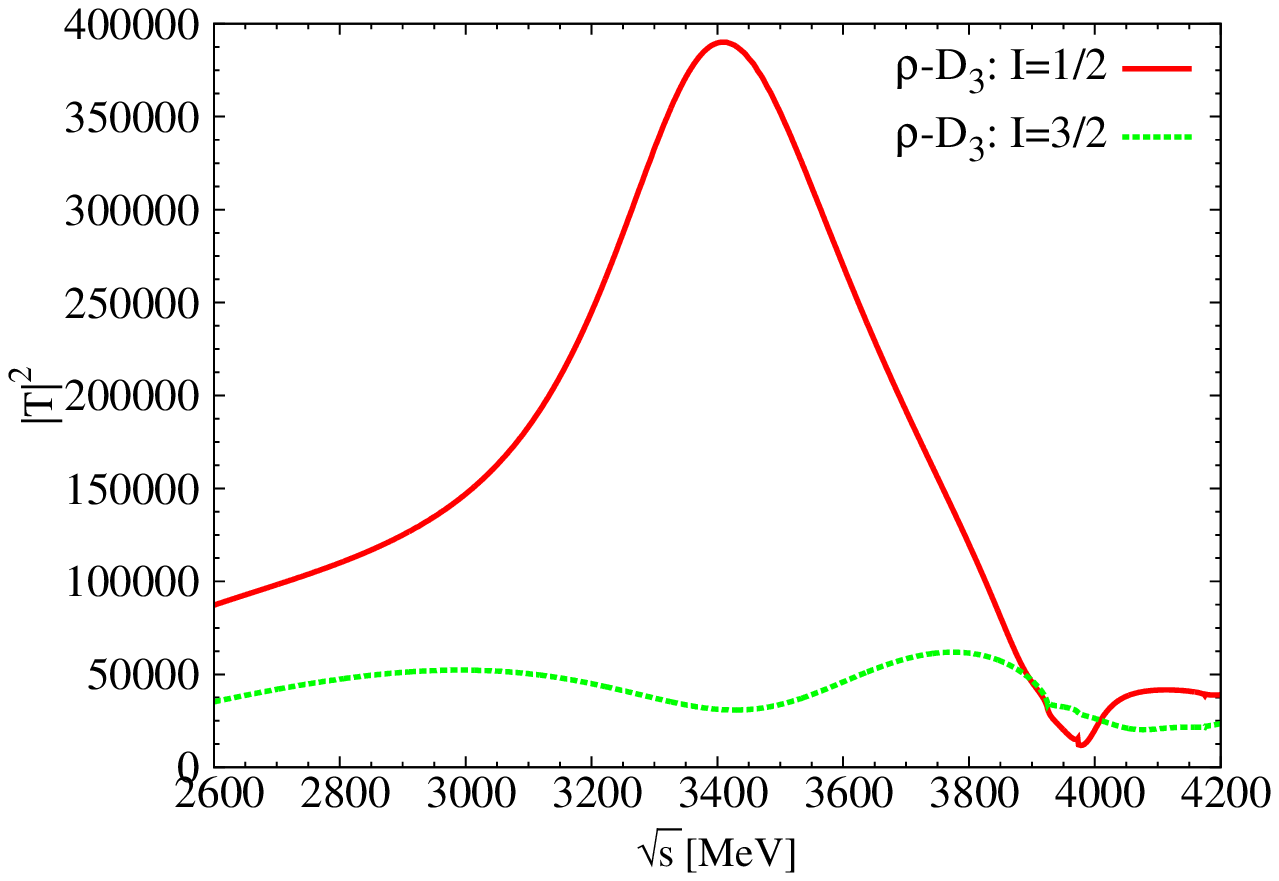}
\caption{Modulus squared of the scattering amplitudes of $T_{D-f_4}$ (left) and $T_{\rho-D_3}$ (right).}
\label{fig:t5Df4rD3}
\end{figure}

Our results are shown in Fig. \ref{fig:t5Df4rD3} for the two cases of the five-body interactions. From these results, there is a very narrow resonant peak around the energy $3732\mev$ with a width of about $9\mev$ in the left panel of Fig. \ref{fig:t5Df4rD3} for the results of $|T_{D-f_4}^{I=1/2}|^2$. By contrast, a wide resonant structure is found in the right panel of Fig. \ref{fig:t5Df4rD3} for the results of $|T_{\rho-D_3}^{I=1/2}|^2$, of which the mass is abut $3412\mev$ and the width $571\mev$. Besides, it seems to be another wide resonant structure at the region of $3774\mev$ with a width of $522\mev$ for the results of $|T_{\rho-D_3}^{I=3/2}|^2$ on the right of Fig. \ref{fig:t5Df4rD3}. Note that, the states of $D-f_4$ (with isopin $I=1/2$) and $\rho-D_3$ (with $I=3/2$) seems to be less bound, of which the binding energy is about $150\mev$ for both of them, and the one of $\rho-D_3$ with $I=1/2$ is bound up to about $500\mev$. Some predicted charm states in the energy region of $3296\sim 3843\mev$ were found in Refs. \cite{Godfrey:2015dva,Liu:2016efm,Lu:2014zua,Matsuki:2016hzk} with the quark models, which cover the range of our predicted masses. Since no evidence in the PDG \cite{pdg2016} for the findings in our model, three new $D_4$ states can be predicted with some uncertainties, two isopin $I=1/2$ states and one $I=3/2$ state, a narrow $D_4 (3730)$ state, a wide $D_4 (3410)$ state and another wide $D_4 (3770)$ state, respectively.

\subsection{Six-body interactions}

For the six-body interactions, analogously to the five-body interactions with a new predicted $D_3$ state in the four-body interactions of Subsec. \ref{fourb}, there are also two options of the clusters, seen in Table \ref{tab:cases}, the states of $f_4$ and $D_3$. Therefore, we let the state $D_1$ or $f_2$ collide with them, having: (i) particle $3=D_1$, cluster $R=f_4$ ($1=f_2,\;2=f_2$), or (ii) $3=f_2$, $R=D_3$ ($1=f_2,\;2=D_1$). Since the isospins of $I_{f_2} = I_{f_4} = 0$ and $I_{D_1} = I_{D_3} = \frac{1}{2}$, the total isospins of the six-body systems can only be $I_{total} = \frac{1}{2}$. For the first case of $D_1-f_4$ interactions, the essential amplitudes $t_1 = t_2 = t_{D_1 f_2} = T_{D_1-f_2}^{(I=1/2)}$ have been evaluated in the four-body interactions of Subsec. \ref{fourb}. In fact, having two cases for the four-body interactions of $D_1-f_2$, there is another possibilities for the input amplitudes of $t_{D_1 f_2}$, which means that one can choose $t_1 = t_2 = t_{D_1 f_2} = T_{f_2-D_1}^{(I=1/2)}$ too. For the second case of $f_2$ colliding with the $D_3$, one of the input amplitudes $t_1 = t_{f_2 f_2} = T_{f_2-f_2}$ can be evaluated by reproducing the results of Ref. \cite{Roca:2010tf}, and the other one $t_2 = t_{f_2 D_1} = T_{f_2-D_1}$ is done in the four-body interactions of Subsec. \ref{fourb}. 

\begin{figure}
\centering
\includegraphics[scale=0.6]{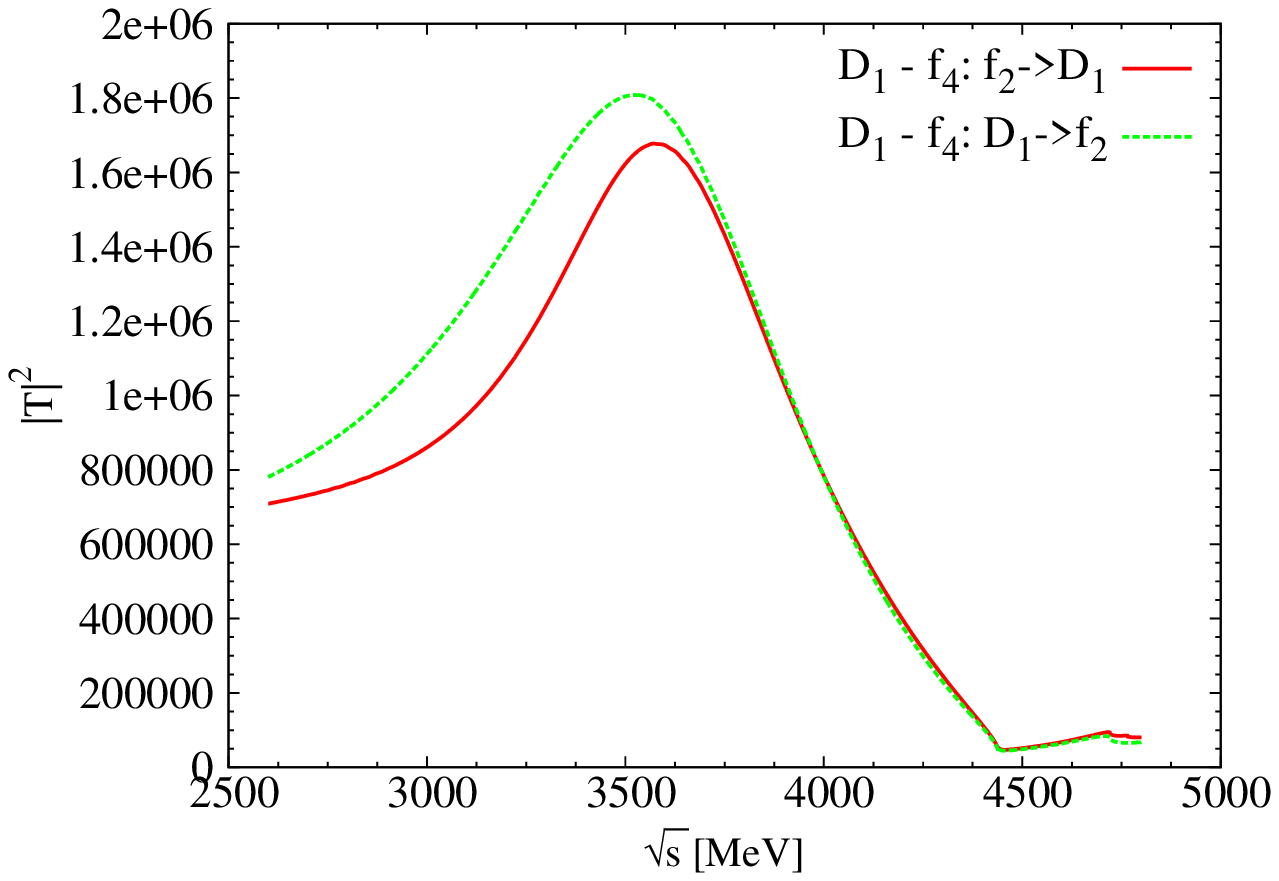}
\includegraphics[scale=0.6]{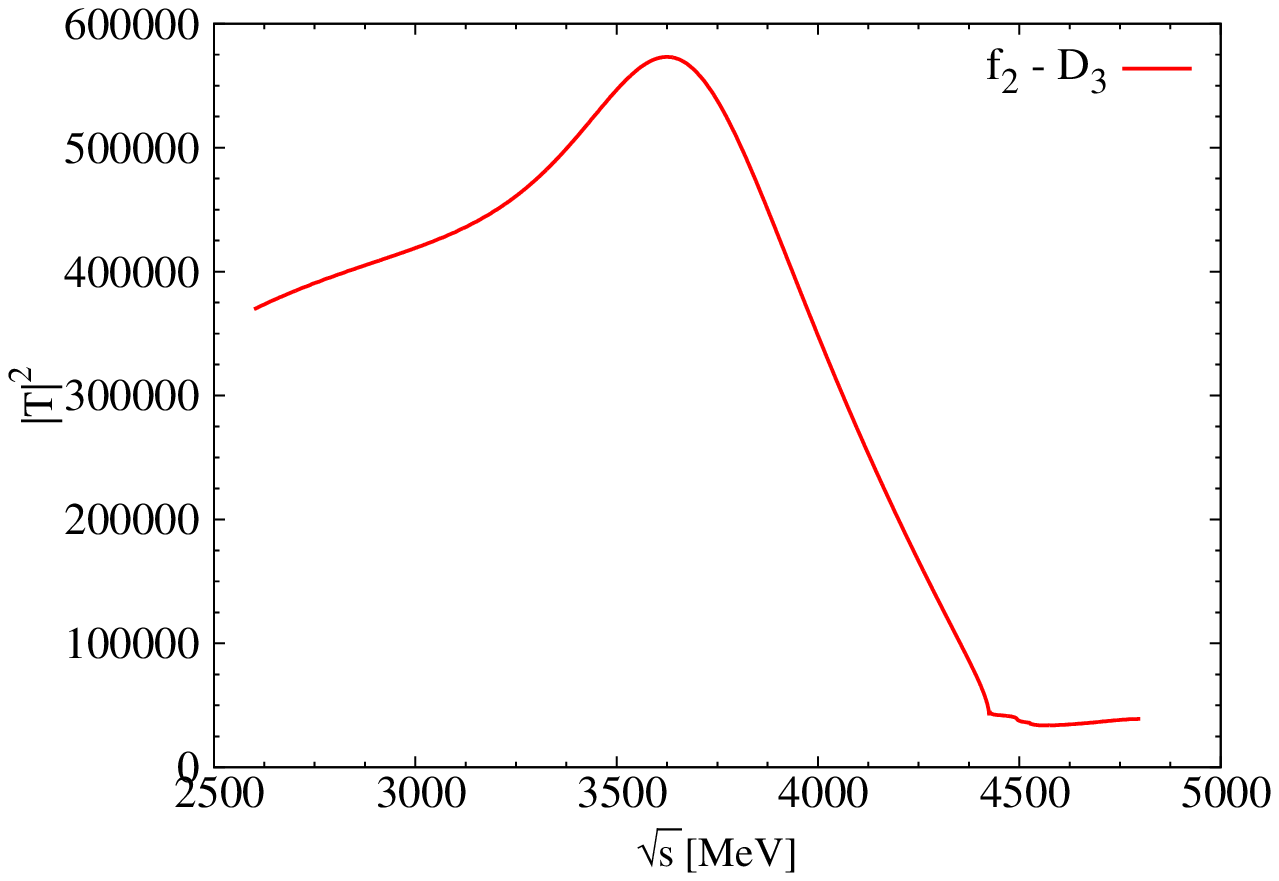}
\caption{Modulus squared of the scattering amplitudes of $T_{D_1-f_4}$ (left) and $T_{f_2-D_3}$ (right).}
\label{fig:t6D1f4f2D3}
\end{figure}

We show our results for the six-body interactions in Fig. \ref{fig:t6D1f4f2D3}, where the results of $|T_{D_1-f_4}^{I=1/2}|^2$ are on the left and the ones of $|T_{f_2-D_3}^{I=1/2}|^2$ on the right. From Fig. \ref{fig:t6D1f4f2D3}, the resonant structures are found in both scattering cases, even though the widths of the peaks are very wide. For the first case, by taking $T_{D_1-f_2}^{(I=1/2)}$ as input amplitudes, we find a peak at the region of about $3569\mev$ with a large width of about $1009\mev$, and taking $T_{f_2-D_1}^{(I=1/2)}$ for the inputs, a resonant peak is located at about $3523\mev$, of which the width is about $1160\mev$. Thus, we can see that the different options for the input amplitudes of $t_{D_1 f_2}$ do not affect much on the results. For the second case of $f_2$ interacting with the $D_3$ state, seen the right panel of Fig. \ref{fig:t6D1f4f2D3}, we can observe another peak at the position of $3625\mev$, of which the width is more than $1200\mev$. Using the quark models, some charm states were predicted in the energy region of $3397\sim 3722\mev$ in Refs. \cite{Godfrey:2015dva,Lu:2014zua}, where our results are within their predictions. For no $D_5$ states in the PDG \cite{pdg2016}, we can predict a new $D_5$ state with more uncertainties from our results, with a mass of about $3523 \sim 3625\mev$ and a very large width about $1009 \sim 1200\mev$ or more (also very large uncertainties in our results), which may be a molecular state of $D_1-f_4$ or $f_2-D_3$ structure.

\section{Conclusions}

Since some $D_J^{(*)}$ states had been found by LHCb experiments, in the present work, we apply the formalism of the fixed-center-approximation to Faddeev equations to investigate the interactions of the $D$-muti-$\rho$ systems, where we search the bound states in the interactions. First, we dynamically reproduce the states of $f_2 (1270)$ and $D_1 (2420)$ in the two-body $\rho\rho$ and $\rho D$ interactions respectively, which are assumed as the fixed center of the three-body interactions. Then, we extrapolate the formalism to the systems of one by one from the three-body systems up to the six-body systems. What we obtain are summarized in Table \ref{tab:summa}, where we have found some bound states in the multi-body interactions. In the three-body interactions, we associate the state $D(3000)^0$ in PDG as a molecular state of $D-f_2$ or $\rho-D_1$ structure with some uncertainties, where we determine its $J^P = 2^-$. There is another possible isospin $I=3/2$ state, $D_2 (3100)$, which has more uncertainties. We find a bound state of $D_3 (3160)$ with $I(J^P) = \frac{1}{2} (3^+)$ in the four-body interactions, which is a molecule of $f_2-D_1$ structure. There are three predictions in the five-body interactions, a narrow $D_4 (3730)$ state with $I(J^P) = \frac{1}{2} (4^-)$, a wide $D_4 (3410)$ state of $I(J^P) = \frac{1}{2} (4^-)$, and another wide $D_4 (3770)$ state but with $I(J^P) = \frac{3}{2} (4^-)$. Finally, with much more uncertainties, we find a $D_5 (3570)$ state with $I(J^P) = \frac{1}{2} (5^+)$ in the six-body interactions, which has a very large width. Hope some predicted states can be searched in the future experiments.

\begin{table}[htb]
   \renewcommand{\arraystretch}{1.7}
   \setlength{\tabcolsep}{0.4cm}
\centering
\caption{Results of the $D$-multi-$\rho$ interactions (units: MeV).}
\label{tab:summa}
\begin{tabular}{cccccc}
\hline\hline
  & Interactions & $I(J^P)$ & Results (mass, width) & PDG & Pridictions  \\
\hline
Three-body
 & $D-f_2\;(\rho\rho)$ & $\frac{1}{2} (2^-)$ & (2997, 14)  &  $D(3000)^0$  & $\cdots$ \\
 & $\rho-D_1\;(\rho D)$ & $\frac{1}{2} (2^-)$ & (2929, 103) &  $D(3000)^0$  & $\cdots$ \\
 & $\rho-D_1\;(\rho D)$ & $\frac{3}{2} (2^-)$ & $(3100, \sim 650)$  &  $\cdots$  & $D_2 (3100)$ \\
\hline
Four-body
  & $f_2-D_1\;(\rho D)$ & $\frac{1}{2} (3^+)$  & (3135, 344) &  $\cdots$ & $D_3 (3160)$ \\
  & $D_1-f_2\;(\rho\rho)$ & $\frac{1}{2} (3^+)$  & (3180, 390) & $\cdots$ & $D_3 (3160)$\\
\hline
Five-body
   & $D-f_4\;(f_2 f_2)$ & $\frac{1}{2} (4^-)$ & (3732, 9) &  $\cdots$  & $D_4 (3730)$  \\
   & $\rho-D_3\;(f_2 D_1)$ & $\frac{1}{2} (4^-)$ & (3412, 571) & $\cdots$  & $D_4 (3410)$  \\
   & $\rho-D_3\;(f_2 D_1)$ & $\frac{3}{2} (4^-)$ & (3774, 522) &  $\cdots$ & $D_4 (3770)$  \\
\hline
Six-body
  & $D_1-f_4\;(f_2 f_2)$ & $\frac{1}{2} (5^+)$ & (3569/3523, 1009/1160) & $\cdots$ & $D_5 (3570)$ \\
  & $f_2-D_3\;(f_2 D_1)$ & $\frac{1}{2} (5^+)$ & (3625, $>1200$) & $\cdots$ & $D_5 (3570)$ \\
\hline\hline
\end{tabular}
\end{table}

\section*{Acknowledgements}

The author thanks Ulf-G. Mei{\ss}ner for the useful comments and discussions, and appreciates Vakhid A. Gani for the useful comments.
This work is supported in part by the DFG and the NSFC through funds provided to the Sino-German CRC~110 ``Symmetries and the Emergence of Structure in QCD''.

\appendix

\section{The box diagram contributions and the convolutions of the loop functions}
\label{appa}

As done in Ref. \cite{Molina:2008jw}, the contributions of the box diagrams should be considered for the corrections of the potentials $V^I$. The main contribution is $\pi\pi$-box diagram for $\rho\rho$ interactions, written
\begin{equation} 
_{\rho\rho}V_{box (\pi\pi)}^{(I=0,S=2)} (s) = 8\tilde{V}^{(\pi\pi)},
\end{equation}
where $\tilde{V}^{(\pi\pi)}$ is given by
\begin{equation}
\begin{split}
\tilde{V}^{(\pi\pi)} (s) =& \frac{32 g^4}{15\pi^2} \int_0^{q'_{max}} d\vec{q} \,\vec{q}\;^6 [10\omega^2 - (k_3^0)^2] \frac{1}{\omega^3} \Big( \frac{1}{k_1^0 + 2\omega} \Big)^2 \frac{1}{P^0 + 2\omega} \\
&\times \frac{1}{k_1^0 + \frac{\Gamma}{4} - 2\omega + i\epsilon} \frac{1}{k_1^0 - \frac{\Gamma}{4} - 2\omega + i\epsilon} \frac{1}{P^0 - 2\omega + i\epsilon} F(\vec{q}\;)^4,
\end{split}
\end{equation}
where the energies $\omega = \sqrt{\vec{q}\;^2+m_\pi^2}$, $\sqrt{s} = P^0 = k_1^0 + k_2^0$, and a cut off of $q'_{max}=1200\mev$ is chosen in the integration, which is within a natural size \cite{Oller:2000fj}. Beside, $F(\vec{q}\;)$ is a form factor for an off-shell pion in each vertex, in the case of $\pi\pi$-box, taken
\begin{equation} 
F(\vec{q}\;) = \frac{\Lambda^2-m_\pi^2}{\Lambda^2+\vec{q}\;^2},
\end{equation}
with a scale of $\Lambda=1300\mev$. 

Next, we discuss how to do the convolutions for the the loop functions. The elements of the matrix $G^I$ in Eq. \eqref{eq:bse} are given by two meson loop functions, having
\begin{equation} 
G_{ii}^I (s,m_1,m_2) = i \int \frac{d^4 q}{(2\pi)^4} \frac{1}{q^2 - m_1^2 + i\epsilon} \frac{1}{(P-q)^2 - m_2^2 + i\epsilon},
\label{eq:loop}
\end{equation}
which can be regularized by the cut off method, obtained
\begin{equation} 
G_{ii}^I (s,m_1,m_2) = \int_0^{q_{max}} \frac{\vec{q}\;^2 d|\vec{q}|}{(2\pi)^2} \frac{\omega_1 + \omega_2}{\omega_1 \omega_2 [(P^0)^2 - (\omega_1 + \omega_2)^2 + i\epsilon]},\label{eq:cutoff}
\end{equation}
where the energies $\omega_i$ are defined as the ones in Eqs. \eqref{eq:formfactor} and \eqref{eq:formfactorN}, the centre-of-mass energy $(P^0)^2=s$, and we use a cutoff of $q_{max}=875\mev$ for the $\rho\rho$ interactions.

Considering a finite width of the vector meson in the loop function, the effects of the propagation of unstable particles are taken into account in terms of the Lehmann representation, which is formulated by the dispersion relation with its imaginary part, written
\begin{equation}
F(s)= \int_{ s_{ {\rm th} }}^{\infty} ds'
  \left( - \frac{1}{\pi} \right)
  \frac{\Ima F(s')}{s - s' + i\epsilon} \; ,
\label{eq:disp}
\end{equation}
where $s_{{\rm th}}$ is the square of the threshold energy, and the spectral function is taken as
\begin{equation}
\Ima F(s') = \Ima \left\{ \frac{1}{s' - M_V^2 + iM_V \Gamma_V} \right\} \; ,
\end{equation}
with $M_V$ and $\Gamma_V$ the mass and the width of the vector meson which can be taken as their physical value in most of cases. Thus, using the Lehmann representation for each $\rho$ meson in the $\rho \rho$ channel, we can take into account the convolutions for the loop functions with the $\rho$ mass distributions, given by
\begin{equation}
\begin{split}
\tilde{G}_{\rho\rho}(s) =& \frac{1}{N^2} \int_{(m_\rho-2\Gamma_\rho)^2}^{(m_\rho+2\Gamma_\rho)^2} d\tilde{m}_1^2 \Big( -\frac{1}{\pi} \Big) \Ima \frac{1}{\tilde{m}_1^2 - m_\rho^2 + i \tilde{m}_1 \Gamma(\tilde{m}_1)}\\
&\times \int_{(m_\rho-2\Gamma_\rho)^2}^{(m_\rho+2\Gamma_\rho)^2} d\tilde{m}_2^2 \Big( -\frac{1}{\pi} \Big) \Ima \frac{1}{\tilde{m}_2^2 - m_\rho^2 + i \tilde{m}_2 \Gamma(\tilde{m}_2)} G_{\rho\rho}(s,\tilde{m}_1,\tilde{m}_2),
\end{split}
\end{equation}
with the normalization factor as
\begin{align}
&N= \int_{(m_\rho-2\Gamma_\rho)^2}^{(m_\rho+2\Gamma_\rho)^2} d\tilde{m}_1^2 \Big( -\frac{1}{\pi} \Big) \Ima \frac{1}{\tilde{m}_1^2 - m_\rho^2 + i \tilde{m}_1 \Gamma(\tilde{m}_1)},\\
&\Gamma(\tilde{m}_1)= \Gamma_\rho \Big( \frac{\tilde{m}_1^2 - 4m_\pi^2}{m_\rho^2 - 4m_\pi^2} \Big)^{3/2},
\end{align}
where $\Gamma_\rho = 146.2\mev$ and $G_{\rho\rho} (s,\tilde{m}_1,\tilde{m}_2)$, is given by Eq. \eqref{eq:cutoff}. Note that, we have taken a energy dependent decay width $\Gamma(\tilde{m}_1)$ for the large decay width of the $\rho$ meson.

For the case of the $\rho D$ interactions, seen the Subsec. \ref{subsecRD}, we use the explicit expression for the loop function with the dimensional regularization to  Eq.~\eqref{eq:loop}, written
\begin{eqnarray}
G_{ii}(s,M_i,m_i) &=& \frac{1}{16\pi^2} \left\{ a(\mu) + {\rm ln} \frac{M_i^2}{\mu^2} + \frac{m_i^2 - M_i^2 + s}{2s} {\rm ln}\frac{m_i^2}{M_i^2} \right. \nonumber \\
  && + \frac{q_i}{\sqrt{s}} \left[ {\rm ln} (s - (M_i^2 - m_i^2) + 2q_i \sqrt{s}) \right. \nonumber \\
  && + {\rm ln} (s + (M_i^2 - m_i^2) + 2q_i \sqrt{s}) \nonumber \\
  && - {\rm ln} (-s + (M_i^2 - m_i^2) + 2q_i \sqrt{s}) \nonumber \\
  && \left. \left. - {\rm ln} (-s - (M_i^2 - m_i^2) + 2q_i \sqrt{s}) \right] \right\} ,
\label{eq:Gdr}
\end{eqnarray}
where $a(\mu)$ is a subtraction constant related to a regularization scale of $\mu$, taking $\mu = 1500 \mev$ and $a(\mu) = -1.55$ from the fits of Ref. \cite{Gamermann:2007fi}, $M_i$ and $m_i$ are the masses of the vector and pseudo-scalar mesons respectively, and the momentum at the center of mass frame, $q_i$, is given by 
\begin{equation}
q_i = \frac{\sqrt{[s- (M_i-m_i)^2][s - (M_i+m_i)^2)] }}{ 2\sqrt{s} }.
\end{equation}

Using the Lehmann representation of Eq.~\eqref{eq:disp}, we can evaluate the convolution for the loop function which only has one vector meson of large width in the channel $i$,
\begin{eqnarray}
\tilde{G}_{ii}(s,M_i,m_i) &=& \frac{1}{N_i} \int_{(M_i - 2 \Gamma_i)^2}^{(M_i + 2\Gamma_i)^2} ds' G_{ii} (s,\sqrt{s'},m_i) \nonumber \\
  && \times  \left(-\frac{1}{\pi} \right) {\rm Im} \left\{\frac{1}{s' - M_i^2 + i M_i \Gamma_i} \right\} ,
  \end{eqnarray}
where $G_{ii}$ is given by Eq.~\eqref{eq:Gdr}, and the normalization factor for the $i$th component
\begin{equation}
N_i = \int_{(M_i - 2 \Gamma_i)^2}^{(M_i + 2\Gamma_i)^2} ds' \times \left( - \frac{1}{\pi} \right) 
{\rm Im} \left\{ \frac{1}{s' - M_i^2 + iM_i \Gamma_i} \right\} , 
\end{equation}
with $\Gamma_i$ the width of the vector meson.

\end{document}